\pdfoutput=1
\documentclass[11pt]{article}

\usepackage[preprint]{acl}

\usepackage{times}
\usepackage{latexsym}

\usepackage[T1]{fontenc}
\usepackage[utf8]{inputenc}

\usepackage{microtype}

\usepackage{inconsolata}

\usepackage{caption}
\usepackage{graphicx}
\usepackage{float} 
\usepackage{subcaption}
\usepackage{amsmath}
\usepackage{amssymb}
\usepackage{hyperref}
\usepackage{url}
\usepackage{hyperref}
\usepackage{booktabs}
\usepackage{enumitem}
\usepackage{multirow}
\usepackage{algorithm2e}
\SetKwComment{Comment}{/* }{ */}
\RestyleAlgo{ruled}

\title{Robust Singing Voice Transcription Serves Synthesis}

\author{
Ruiqi Li, 
Yu Zhang, 
Yongqi Wang, 
Zhiqing Hong,  
Rongjie Huang,
Zhou Zhao\thanks{Corresponding author}
\\
Zhejiang University \\
\texttt{\{ruiqili,yuzhang34,rongjiehuang,zhaozhou\}@zju.edu.cn}
}

\begin{document}
\maketitle
\begin{abstract}

% version 1
% Note-level automatic singing voice transcription (AST) aims to convert singing voice recordings into note sequences, which makes automatic annotation of singing datasets possible, ultimately serving singing voice synthesis (SVS). 
% However, applying current AST approaches to automatic annotation still faces challenges, such as insufficient accuracy and inadequate robustness. 
% This paper presents the first robust AST model that serves SVS. 
% Based on a multi-scale architecture, our model effectively captures coarse-grained note information and ensures fine-grained frame-level segmentation. 
% An attention-based pitch decoder is devised to predict pitch values resiliently. 
% To explore real-world scenarios, we establish a comprehensive annotation-and-training pipeline for an SVS task. 
% Experimental findings reveal that the proposed model achieves state-of-the-art transcription accuracy with either clean or noisy inputs. 
% Furthermore, the SVS model, trained with expanded and automatically annotated datasets, outperforms its baseline, affirming the capability for practical application. Audio samples are available at \href{https://rosvot.github.io}{rosvot.github.io}.

Note-level Automatic Singing Voice Transcription (AST) converts singing recordings into note sequences, facilitating the automatic annotation of singing datasets for Singing Voice Synthesis (SVS) applications. 
Current AST methods, however, struggle with accuracy and robustness when used for practical annotation. 
This paper presents ROSVOT, the first robust AST model that serves SVS, incorporating a multi-scale framework that effectively captures coarse-grained note information and ensures fine-grained frame-level segmentation, coupled with an attention-based pitch decoder for reliable pitch prediction.
We also established a comprehensive annotation-and-training pipeline for SVS to test the model in real-world settings. 
Experimental findings reveal that ROSVOT achieves state-of-the-art transcription accuracy with either clean or noisy inputs. 
Moreover, when trained on enlarged, automatically annotated datasets, the SVS model outperforms its baseline, affirming the capability for practical application.
Audio samples are available at \href{https://rosvot.github.io}{https://rosvot.github.io}. Codes can be found at \href{https://github.com/RickyL-2000/ROSVOT}{https://github.com/RickyL-2000/ROSVOT}.

\end{abstract}

% This paper introduces a pioneering and robust approach to automatic singing voice transcription (AST) specifically tailored for singing voice synthesis (SVS) applications. Our model, based on a multi-scale architecture, adeptly captures coarse-grained note information while ensuring fine-grained frame-level localization. The incorporation of an attention-based pitch decoder enhances the robust prediction of pitch values. To explore real-world applications, we establish a comprehensive annotation-and-training pipeline for an SVS task. Experimental findings reveal that the proposed model attains state-of-the-art transcription accuracy under various conditions, exhibiting resilience to both clean and noisy inputs. Furthermore, the SVS model, trained with expanded and automatically annotated datasets, surpasses baseline performance, affirming its practical applicability.

\section{Introduction}
Note-level automatic singing voice transcription (AST) refers to converting a singing voice recording into a sequence of note events, including note pitches, onsets, and offsets \cite{mauch2015computer, hsu2021vocano, wang2022musicyolo, yong2023phoneme}. As part of the music information retrieval (MIR) task, AST is widely used in professional music production and post-production tuning. 
% With the recent advancements in singing voice synthesis (SVS), there is an increasing demand for annotated data,  AST has also demonstrated the potential for automated annotation. 
With the recent advancements of singing voice synthesis (SVS) \cite{Liu_Li_Ren_Chen_Zhao_2022, huang2022singgan, 9747664, he2023rmssinger}, there is a growing demand for annotated data, while AST methods just demonstrate the potential for automatic annotation. 

% As shown in \autoref{fig:ast}, a common data collection pipeline for SVS consists of two stages: a) phoneme/word annotation and b) note annotation, where the former can be achieved by utilizing automatic speech recognition (ASR) approaches and forced alignment tools, such as MFA \cite{mcauliffe2017montreal}. The second stage, however, is far from reaching a fully automatic level. Arduous manual annotation hinders large-scale data collection.

Note transcription from singing voices is particularly difficult than from musical instruments, as the pitch component of human voices is highly dynamic. When singing, people articulate words, leading to unstable pitches and blurry note boundaries. For instance, if a word starts with a voiceless consonant, the pitch onset may be slightly delayed. 
Also, singing techniques like vibrato and appoggiatura further complicate boundary localization.
% not to mention melisma – the singing of a single syllable while transitioning between different notes in succession.
% \citet{yong2023phoneme} introduces additional phoneme-related information to improve boundary detection.

\vspace{-8pt}

\begin{figure}[!htbp]
    \centering
    \includegraphics[width=0.45\textwidth]{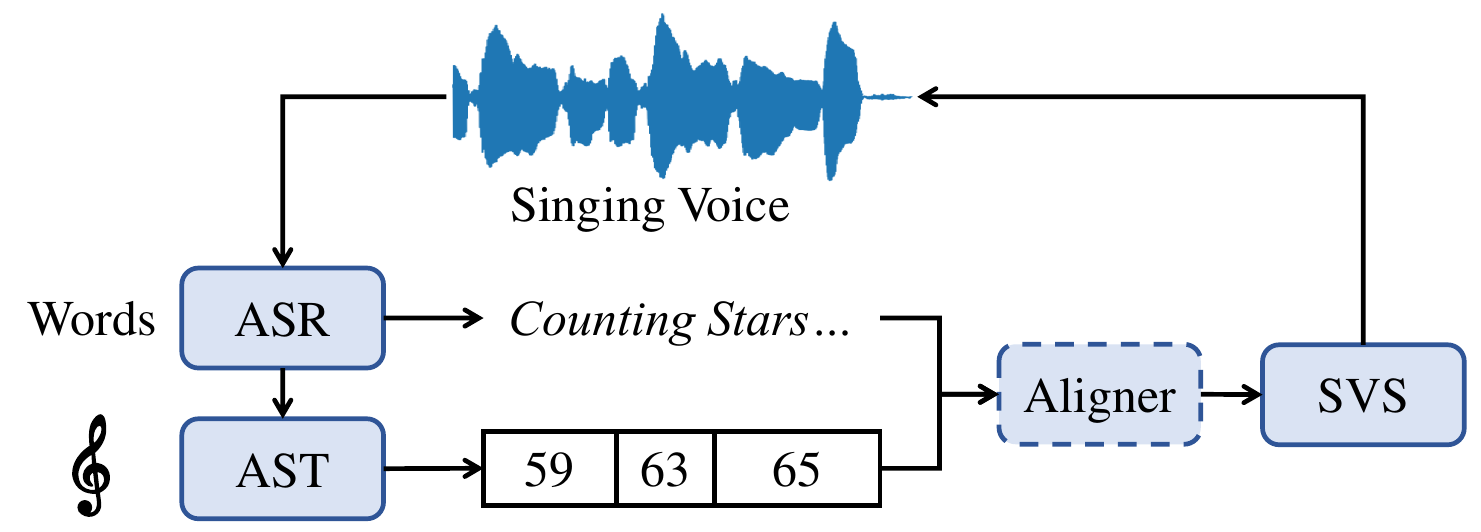}
    \setlength{\belowcaptionskip}{-0.3cm}
    \caption{
        AST and ASR systems serve SVS.
    }
    \label{fig:ast}
\end{figure}

An AST task is mainly decomposed into two steps: note segmentation and pitch estimation. The first step predicts boundaries, or onset and offset of each note, which is always implemented as classification \cite{hsu2021vocano, yong2023phoneme} or object detection \cite{wang2022musicyolo} tasks. For pitch estimation, previous works primarily adopt weighted median or average operations on F0 values. 

Despite previous accomplishments, there is no AST model that, to our knowledge, achieves a complete annotation pipeline for training an SVS model.
Applying AST approaches to automated annotation for SVS tasks still faces several challenges:

\vspace{-4pt}

\begin{itemize}[leftmargin=*]
    \setlength{\itemsep}{-4pt}
    \item \textbf{Insufficient accuracy.} Despite numerous efforts to improve accuracy, the performance is still insufficient for automatic annotation. Currently, AST results serve merely as a preliminary guide, necessitating additional manual refinement for actual application \cite{zhang2022m4singer}. 
    \item \textbf{Asynchronization between notes and texts.} SVS models often require text-note synchronized annotation. 
    % A common data collection pipeline consists of two stages: a) phoneme/word annotation and b) note annotation, where the former can be achieved by utilizing forced alignment tools, such as MFA \cite{mcauliffe2017montreal}, or manual annotation. 
    Currently, transcribing singing voices without the supervision of word/phoneme boundaries requires additional post-processing for alignment, introducing accumulative errors. 
    % However, word boundary information is not hard to obtain, either from stage (a) or an extra pre-trained word boundary predictor. 
    % We can use word boundary information to regulate note segmentation. 
    \item \textbf{Inadequate robustness.} Web crawling is a popular method for data collection \cite{ren2020deepsinger},
    but the quality varies.
    % However, the quality of audio obtained through the internet varies. 
    % and some even require vocal accompaniment separation first, making the audio slightly distorted and noisy. 
    AST methods are vulnerable to noise as sound artifacts tend to disrupt boundary localization and pitch perception. 
\end{itemize}

\vspace{-4pt}

In this paper, we present ROSVOT, a \textbf{RO}bust \textbf{S}inging \textbf{VO}ice \textbf{T}ranscription model that ultimately serves SVS. The note boundary prediction is formulated as one-dimensional semantic segmentation, and an attention-based decoder is employed for pitch prediction.
To achieve both coarse-grained semantic modeling and fine-grained frame-level segmentation, we devise a multi-scale architecture by integrating Conformer \cite{gulati2020conformer} and U-Net \cite{ronneberger2015u}. 
% We leverage Conformer to model the coarse-grained note event-related features from sung representations. To alleviate the long-sequence computational burden of the Conformer and reconstruct high-resolution spectrograms, we utilize a U-Net backbone to "wrap" the Conformer, making it focus on downsampled representations. 
Moreover, the model incorporates word boundaries to guide the segmentation process. We randomly mix the input waveforms with MUSAN \cite{musan2015} noise to simulate a noisy environment, forming a bottleneck and bolstering denoising capabilities.

To demonstrate the potential of ROSVOT in practical annotation applications, we conduct extensive experiments on a comprehensive annotation-and-training pipeline on an SVS task, simulating real-world scenarios. We choose and slightly modify RMSSinger \cite{he2023rmssinger}, one of the state-of-the-art SVS models, to be the singing acoustic model. Experiments show that the SVS model trained with pure transcribed annotations achieves 91\% of the pitch accuracy compared to manually annotated data, without loss of overall quality. 
% Furthermore, we investigate and train the proposed model in a low-resource setting and introduce external self-supervised features to alleviate data scarcity. 
We also explore the generalization performance on cross-lingual tasks, where we use ROSVOT trained with Mandarin corpora to annotate an English corpus, which is then used to train an SVS model.
Our contributions are summarized as follows:

% \vspace{-2pt}
% \begin{itemize}[leftmargin=*]
%     \setlength{\itemsep}{0pt}
%     \item We propose the first robust AST model towards singing voice synthesis. 
%     \item We design a multi-scale AST model, which achieves state-of-the-art transcription accuracy under either high-fidelity recording or noisy environment. 
%     \item We construct an annotation-and-training pipeline to investigate the effect of automatically transcribed annotations on SVS tasks. Experiments demonstrate that the proposed model has the capability for practical application.
%     % \item We investigate low-resource scenarios and explore cross-lingual generalization capabilities.
%     \item We explore the cross-lingual generalization capabilities of automatic annotation.
% \end{itemize}
% \vspace{-2pt}

\vspace{-8pt}

\begin{itemize}[leftmargin=*]
    \setlength{\itemsep}{-4pt}
    \item We propose ROSVOT, the first robust AST model that serves SVS, which achieves state-of-the-art transcription accuracy under either clean or noisy environments. 
    \item We construct a comprehensive annotation-and-training pipeline to investigate the effect of automatically transcribed annotations on SVS tasks. 
    \item The proposed multi-scale model outperforms the previous best published method by 17\% relative improvement on pitch transcription, and by 23\% with noisy inputs. 
    \item By incorporating automatically annotated large-scale datasets, we demonstrate ROSVOT's capability of practical application and the opportunity to alleviate data scarcity in SVS. 
    \item We explore the cross-lingual generalization capabilities of ROSVOT. 
\end{itemize}
\vspace{-4pt}

\section{Related Works}

\subsection{Automatic Singing Voice Transcription}

% AST is an automatic music transcription (AMT) task \cite{bhattarai2023comprehensive}, belonging to the MIR family. 
AST is useful not only in automatic music transcription (AMT) \cite{bhattarai2023comprehensive}, but also a promising task for audio language models \cite{yang2023uniaudio} and speech-singing interaction modeling \cite{li2023alignsts}. TONY \cite{mauch2015computer} predicts note events by applying hidden Markov models (HMM) on extracted pitch contours. VOCANO \cite{fu2019hierarchical, hsu2021vocano} considers the note boundary prediction as a hierarchical classification task.
and leverages a hand-crafted signal representation for feature engineering. 
MusicYOLO \cite{wang2022musicyolo} adopts object detection methods from image processing to localize the onset and offset positions. 
% Despite their success, the overall accuracy is still insufficient for practical application. 
Considering the linguistic characteristic of singing voices, \citet{yong2023phoneme} introduces extra phonetic posteriorgram (PPG) information to improve accuracy. 
% , because special articulation patterns may influence the note formation process. 
% However, we believe that PPG conditions introduce unnecessary information and disturbs the information bottleneck. 
However, a PPG extractor requires an extra training process and makes the AST model difficult to generalize across languages.
\citet{gu2023deep} attempts to achieve robust transcription through self-supervised pre-training and multimodal injection. 

\subsection{Singing Voice Synthesis}

Recently, there has been notable progress in the field of SVS. HifiSinger \cite{chen2020hifisinger} and WeSinger \cite{zhang2022wesinger} employ GAN-based networks for high-quality synthesis. \cite{Liu_Li_Ren_Chen_Zhao_2022} introduces a shallow diffusion mechanism to address over-smoothness issues in the general Text-to-Speech (TTS) field. Taking inspiration from VITS \cite{kim2021vits}, VISinger \cite{9747664} constructs an end-to-end architecture. To achieve singer generalization, NaturalSpeech 2 \cite{shen2023naturalspeech} and StyleSinger \cite{zhang2023stylesinger} utilize a reference voice clip for timbre and style extraction. To bridge the gap between realistic musical scores and MIDI annotations, RMSSinger \cite{he2023rmssinger} proposes word-level modeling with a diffusion-based pitch modeling approach. Open-source singing voice corpora also boost the development of SVS \cite{huang2021multi, zhang2022m4singer, wang2022opencpop}. However, the quantity of annotated singing voice corpora is still small compared to speech, while note annotations of some corpora are even unavailable. 

\section{Method}

\subsection{Problem Formulation}

% As stated before, an AST task is often decomposed into a note segmentation step and a pitch estimation step. 
In the note segmentation step, the model predicts onset/offset states at each timestep $t$, where $t \in [1, T]$ and $T$ is the temporal length of the spectrogram. Without loss of generality, we introduce silence notes to connect each note in the entire sequence end-to-end, replacing the onset/offset tuples by a single note boundary notation sequence $\boldsymbol{y}_{\text{bd}} = [y_{\text{bd}}^1, y_{\text{bd}}^2, ..., y_{\text{bd}}^T]$, where $y_{\text{bd}}^t = 1$ if the state is boundary at timestep $t$ and 0 is not. The silence note has a pitch value of 0. 
% Notice that the count of note boundaries should be less than the number of notes by 1, or formally: 
Notice that $\sum \boldsymbol{y}_{\text{bd}} = \text{len}(\boldsymbol{p}) - 1$, where $\boldsymbol{p} = [p^1, p^2, ..., p^{L}]$ is the pitch value sequence, $L$ is the total number of notes, and $\text{len}(\cdot)$ computes lengths of sequences. Therefore, the first step can be treated as semantic segmentation, predicting a binary-label sequence. The second step is to predict the pitch sequence $\boldsymbol{p}$.

\subsection{Overview}

    \begin{figure*}[htbp]
        \centering
        \includegraphics[width=0.9\textwidth]{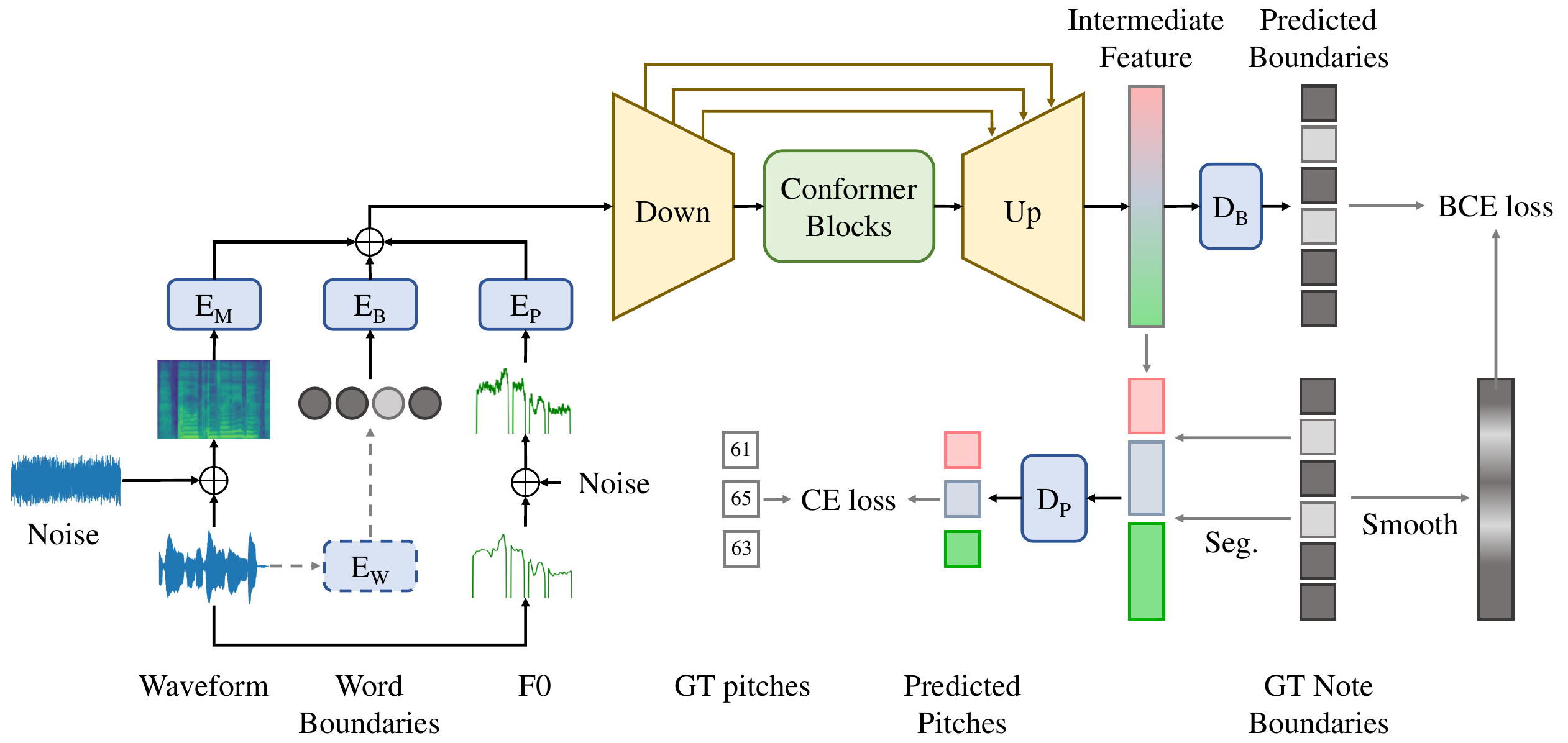}
        \setlength{\abovecaptionskip}{0.1cm}
        \setlength{\belowcaptionskip}{-0.5cm}
        \caption{
            The overall architecture. $\text{E}_{\text{M}}$, $\text{E}_{\text{B}}$, and $\text{E}_{\text{P}}$ represent encoders of Mel-spectrogram, word boundaries, and F0 contour input. $\text{D}_{\text{B}}$ and $\text{D}_{\text{P}}$ stand for decoders of note boundaries and pitches. The "Down" and "Up" parts denote the encoder and decoder of the U-Net backbone. The "Seg." and "Smooth" notations indicate temporal segmentation and label smoothing operations. $E_W$ indicates an optional extractor used to provide word boundaries. 
        }
        \label{fig:main-model}
    \end{figure*}

As shown in \autoref{fig:ast}, a common data collection pipeline for SVS consists of two stages: a) phoneme/word annotation and b) note annotation, where the former can be achieved by utilizing automatic speech recognition (ASR) approaches and forced alignment tools, such as MFA \cite{mcauliffe2017montreal}. The second stage, however, is far from reaching a fully automatic level. Arduous manual annotation hinders large-scale data collection. A high-precision and robust annotator is required.

Note segmentation is a multi-scale classification task, in that the note events are coarse-grained while the predicted boundary sequence $\boldsymbol{y}_{\text{bd}}$ is fine-grained. 
Therefore, we construct a multi-scale model, combining a U-Net backbone and a downsampled Conformer, as illustrated in \autoref{fig:main-model}. 
The model takes Mel-spectrograms, F0 contours, and word boundaries as inputs. 
To improve robustness, we train the model under noisy environments and apply various data augmentation operations. 
For pitch prediction, we adopt an attention-based method to obtain dynamic temporal weights and perform weighted averages. The note segmentation part and the pitch prediction part are trained jointly to acquire optimal results.

\subsection{Data Augmentation}

\subsubsection{Label Smoothing}

The exact temporal positions of note onset and offset are difficult to demarcate on a microscopic scale, because transitions between notes are continuous and smooth. Therefore, label smoothing is a popular strategy in AST tasks \cite{hsu2021vocano, yong2023phoneme}. Also, soft labels carry more information than hard labels, such as the desired confidence for the model. Specifically, we apply temporal convolution operation between the label sequence $\boldsymbol{y}_{\text{bd}}$ and a Gaussian filter $\mathcal{G}[n]$:
\begin{align}
    & \mathcal{G}[n] = \left\{
                         \begin{array}{lr}
                             \frac{1}{\sqrt{2\pi}\sigma}e^{-\frac{\tau^2}{2\sigma^2}}, & \text{if } |n| \leq \lfloor\frac{W_{\mathcal{G}}}{2}\rfloor \\
                             0, & \text{otherwise}
                         \end{array}
                      \right. \\ 
    & \widetilde{\boldsymbol{y}}_{\text{bd}} = \boldsymbol{y}_{\text{bd}} * \left( \frac{\mathcal{G}[n]}{\text{max}(\mathcal{G}[n])} \right)
\end{align}
where the filter $\mathcal{G}[\tau]$ is normalized before convolution, so the middle of each soft label remains 1. $W_{\mathcal{G}}$ indicates the window length of the filter.

\subsubsection{Noise}

We mix realistic noise signals with waveforms before extracting spectrograms. MUSAN noise corpus is utilized to randomly incorporate the interference. MUSAN corpus consists of a variety of noises, such as babble, music, noise, and speech. The intensity of incorporated noise is randomly adjusted according to a signal-to-noise ratio (SNR) interval of $[6, 20]$. The noise signal $\boldsymbol{\eta}$ is repeated or chunked to meet the length of each training sample. In the training stage, we conduct noise mixing followed by on-the-fly extraction of Mel-spectrograms:
\begin{align}
    & \widetilde{\boldsymbol{y}} = \boldsymbol{y} + \boldsymbol{\eta} \times \frac{\text{RMS}(\boldsymbol{y} / 10^{(\text{SNR}/20)})}{\text{RMS}(\boldsymbol{\eta})} \\
    & \widetilde{X} = \mathcal{F}(\widetilde{\boldsymbol{y}})
\end{align}
where $\mathcal{F}(\cdot)$ is Mel-spectrogram extraction operation, $\text{RMS}(\cdot)$ is root-mean-square operation, and $\widetilde{X}$ is the resulting spectrogram.

In addition to spectrograms, we also add noise to F0 contours and label sequences. Since the model takes F0 contours as input, a clean F0 contour can leak information. 
We simply add Gaussian noise to logarithmic F0 contours and soft labels to improve robustness. 

    \begin{figure}[htbp]
        \centering
        \includegraphics[width=0.45\textwidth]{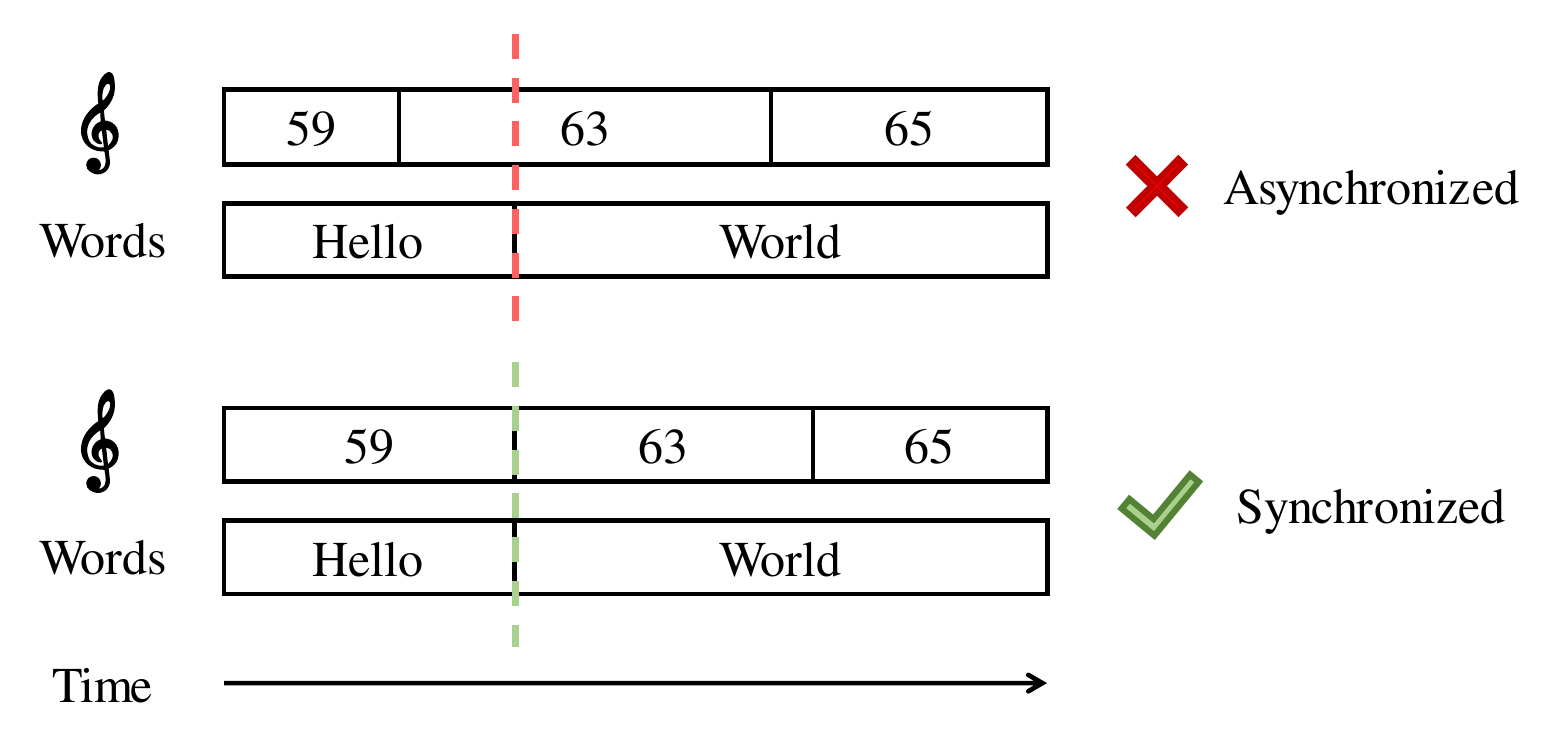}
        \setlength{\belowcaptionskip}{-0.5cm}
        \caption{
            Word-note synchronization.
        }
        \label{fig:sync}
    \end{figure}

\subsection{Word Boundary Condition}

To regulate segmentation results and better suit practical annotation, we incorporate word boundary conditions. The word boundary sequence $\boldsymbol{y}_{\text{wbd}}$ has the same form as note boundaries $\boldsymbol{y}_{\text{bd}}$, involving silence or "NONE" words. The regulation is necessary because, in practical annotation, word sequence and note sequence need to be temporally synchronized, as shown in \autoref{fig:sync}. In other words, the presence of a word boundary at timestep $t$ implies the existence of a note boundary at $t$, but the reverse may not hold true. This is because melisma is a commonly used singing technique. Without regulation, additional post-processing is required to synchronize words and note sequences. 

Since in practice, the note annotation stage follows the phoneme annotation stage, word boundaries should already be obtained through forced alignment tools like MFA. We directly encode word boundaries as an additional condition to ensure word-note synchronization. Moreover, to provide note-only support, we train an extra word boundary extractor $\text{E}_{\text{W}}$ to deal with scenarios like vocal tuning in music industries, where word alignment is unavailable. More details are listed in \autoref{sec:wbd}. 

% We apply three encoders $\text{E}_{\text{M}}$, $\text{E}_{\text{B}}$, $\text{E}_{\text{P}}$ to encode Mel-spectrograms, word boundaries, and F0 contours. The encoders consist of a linear projection or an embedding layer, followed by residual convolution blocks. 

\subsection{Multi-scale Architecture}

% 可以讲讲 word bd 输入, 一堆 encoder

% We leverage Conformer to model the coarse-grained note event-related features from sung representations. To alleviate the long-sequence computational burden of the Conformer and reconstruct high-resolution spectrograms, we utilize a U-Net backbone to "wrap" the Conformer, making it focus on downsampled representations. 

% Note segmentation is a multi-scale classification task, in that the note events are coarse-grained while the predicted boundary sequence $\boldsymbol{y}_{\text{bd}}$ is fine-grained. 

The semantic information of note events is coarse-grained and high-level, while the segmentation result $\boldsymbol{y}_{\text{bd}}$ is fine-grained and frame-level. To tackle this problem, we design a multi-scale model, incorporating multiple feature encoders and a pitch decoder, illustrated in \autoref{fig:main-model}. 

% High-resolution segmentation result is required, otherwise rounding errors occur. Therefore, we adopt a U-Net backbone to downsample representations and guarantee a fine-grained reconstruction. To model the high-level note event-related features, we leverage a Conformer network, one of the most popular ASR models. The U-Net backbone "wraps" the Conformer, making it focus on downsampled features and alleviating long-sequence computational burden. The skip connections improve accurate frame-level localization through multi-scale feature fusion. 
For precise segmentation, high-resolution results are essential to prevent rounding errors. Hence, we employ a U-Net architecture for its ability to downsample representations while ensuring detailed reconstruction. To capture the high-level features associated with note events, we utilize a Conformer network, one of the most popular ASR models. The U-Net architecture envelops the Conformer, directing its focus towards the downsampled features and easing the computational load of processing long sequences. Through the integration of skip connections, our model achieves refined frame-level accuracy by fusing features across multiple scales.

The U-Net backbone's encoder and decoder each comprise $K$ downsampling and upsampling layers, respectively. 
% A downsampling layer consists of a residual convolution block and an average pooling layer with a downsampling rate of 2, resulting in an overall downsampling rate of $2^K$. For an upsampling layer, the input feature is firstly upsampled through a transposed convolution layer, and is then concatenated with the corresponding skipped feature before a final convolution block. 
The downsampling rate is set to 2, and the channel dimension remains the same as input to alleviate overfitting. 
The intermediate part of the backbone is replaced by a 2-layer Conformer block with relative position encoding \cite{dai2019transformer}. The detailed architecture is listed in \autoref{sec:arch}.

\subsection{Decoders and Objectives}

\subsubsection{Note Segmentation}

We adopt a note boundary decoder, denoted as $\text{D}_{\text{B}}$, to transform the output feature $Z$ from the U-Net backbone into logits $\hat{\boldsymbol{y}}_{\text{bd}}$, where $Z \in \mathbb{R}^{T \times C}$ and $C$ is the channel dimension. $\text{D}_{\text{B}}$ is implemented by a single matrix $W_\text{B} \in \mathbb{R}^{C \times 1}$. A binary cross-entropy (BCE) loss $\mathcal{L}_{\text{B}}$ is applied to train note segmentation. The details of loss functions are listed in \autoref{sec:loss}, similarly hereinafter.
% \begin{align}
%     \mathcal{L}_{\text{B}} = & \frac{1}{T} \sum \text{ BCE}(\boldsymbol{y}_{\text{bd}}, \hat{\boldsymbol{y}}_{\text{bd}}) \nonumber \\
%                   = & - \frac{1}{T} \sum_{t=1}^{T} ( y_{\text{bd}}^t \ln(\sigma(\hat{y}_{\text{bd}}^t / T_1)) \nonumber \\
%             & + (1 - y_{\text{bd}}^t) \ln(1 - \sigma(\hat{y}_{\text{bd}}^t / T_1)) )
% \end{align}
% where $T_1$ is the temperature hyperparameter, and $\sigma(\cdot)$ stands for sigmoid function.

It is worth mentioning that in the note segmentation task, there is a significant imbalance between positive and negative samples, with a ratio of approximately 1:500\footnote{Statistically, there are approximately 2.42 note boundaries per second in our datasets.}. 
% In addition, due to the presence of word boundary conditions, the classification difficulty for some boundaries is lower than others. 
Also, the inclusion of word boundary conditions results in varying classification difficulties, with some boundaries being inherently easier to classify than others.
To tackle this imbalance problem, we employ a focal loss \cite{lin2017focal}, $\mathcal{L}_{\text{FC}}$, to focus more on hard samples.
% \begin{align}
%     & p_{\text{t}} = \boldsymbol{y}_{\text{bd}} \sigma(\hat{\boldsymbol{y}}_{\text{bd}}) + (1 - \boldsymbol{y}_{\text{bd}}) (1 - \sigma(\hat{\boldsymbol{y}}_{\text{bd}})) \\
%     & \alpha_{\text{t}} = \alpha \boldsymbol{y}_{\text{bd}} + (1 - \alpha) (1 - \boldsymbol{y}_{\text{bd}}) \\
%     & \mathcal{L}_{\text{FC}} = \frac{1}{T} \sum \alpha_{\text{t}} (1 - p_{\text{t}})^{\gamma} \text{BCE}(\boldsymbol{y}_{\text{bd}}, \hat{\boldsymbol{y}}_{\text{bd}})
% \end{align}
% where $\alpha$ is a hyperparameter controlling weight of positive samples, and $\gamma$ controls balance between easy and hard samples. 

\subsubsection{Pitch Prediction}

For pitch value prediction $\text{D}_{\text{P}}$, we leverage an attention-based weighted average operation to aggregate the fine-grained features, instead of simply applying a weighted median or average. Given the output feature $Z \in \mathbb{R}^{T \times C}$, we obtain an attention weight matrix $S$ through a projection matrix $W_{\text{A}} \in \mathbb{R}^{C \times H}$: $S = \sigma(ZW_{\text{A}})$, 
% \begin{equation}
%     S = \sigma(ZW_{\text{A}})
% \end{equation}
where $S \in \mathbb{R}^{T \times H}$ and $H$ denotes the number of attention heads. Then we perform an outer product operation between each vector of $Z$ and $S$ along the time dimension to obtain a pre-weighted representation: $Z_1^{t} = Z^{t} \otimes S^{t}$ and $Z_1 \in \mathbb{R}^{T \times C \times H}$,
which is further averaged along the head dimension to acquire the weighted representation $Q \in \mathbb{R}^{T \times C}$. In addition, we compute the averaged weights $\boldsymbol{s} \in \mathbb{R}^T$ by averaging along the head dimension.

Subsequently, we use the note boundary sequence $\boldsymbol{y}_{\text{bd}}$ to segment $Q$ along the time axis, resulting in a group sequence $\boldsymbol{G} = [G^1, G^2, ..., G^{L}]$ with length of $L$, number of notes. Each group (or segment) $G^{i}$ contains $l_i$ consecutive vectors: $G^i = [Q^{j+1}, Q^{j+2}, ..., Q^{j+l_i}]$, where $\sum_{i=1}^{L} l_i = T$, $i \in [1, L]$, $j \in [1, T]$, and $y_{\text{bd}}^j = 1$. We also do the same for the averaged weights $\boldsymbol{s}$: $G_s^i = [s^{j+1}, s^{j+2}, ..., s^{j+l_i}]$. For each group, we compute a weighted average $\boldsymbol{z}^i$:
\begin{equation}
    \boldsymbol{z}^i = \frac{ \sum G^i }{ \sum G_s^i } = \frac{ \sum_{k=1}^{l_i} Q^{j+k} }{ \sum_{k=1}^{l_i} s^{j+k} }
\end{equation}

Finally, we multiply $\boldsymbol{z}$ with a matrix $W_O$ to compute the logits: $\hat{\boldsymbol{p}} = \boldsymbol{z} W_{\text{O}}$, where $\boldsymbol{z} \in \mathbb{R}^{L \times C}$ and $W_{\text{O}} \in \mathbb{R}^{C \times P}$. $P$ is the number of pitch categories. A cross-entropy (CE) loss, $\mathcal{L}_{\text{P}}$, is utilized.
% \begin{equation}
%     \mathcal{L}_{\text{P}} = - \frac{1}{L} \sum_{i=1}^{L} \sum_{c=1}^{P} p_c^i \ln \left(  \frac{\exp (\hat{p}_c^i / T_2)}{ \sum_{k=1}^{C} \exp (\hat{p}_c^k / T_2)}  \right)
% \end{equation}
% where $T_2$ is the temperature hyperparameter.

\subsection{Training and Inference Pipeline}

In the training stage, we use ground-truth (GT) note boundaries to segment the intermediate features and optimize the pitch decoder. The overall loss $\mathcal{L} = \lambda_{\text{B}}\mathcal{L}_{\text{B}} + \lambda_{\text{FC}}\mathcal{L}_{\text{FC}} + \lambda_{\text{P}}\mathcal{L}_{\text{P}}$ is controlled by balancing parameters $\lambda_{\text{B}}$, $\lambda_{\text{FC}}$, and $\lambda_{\text{P}}$.

In the inference stage, firstly we compute the boundary probability $\sigma(\hat{\boldsymbol{y}}_{\text{bd}})$ and use a threshold $\mu$ to decide the boundary state. That is, a note boundary exists at time step t if $\sigma(\hat{\boldsymbol{y}}_{\text{bd}}) > \mu$, otherwise, it does not. The predicted results will undergo post-processing to clean up boundaries with excessively small spacing between them. Finally, we segment the intermediate feature $Z$ and decode pitches. 

It is worth mentioning that $\mu$ can control the granularity of generated notes. In other words, a lower $\mu$ may result in more fine-grained and subdivided pitches, while a higher one ignores small fluctuations. This is because a lower $\mu$ allows more boundaries. 

\subsection{Singing Voice Synthesis System}

Once we complete the inference and automatically annotate a dataset, the new datasets are used to train an SVS system to further investigate the practical performance. 
We choose RMSSinger as the singing acoustic model and a pre-trained HiFi-GAN \cite{kong2020hifi} model as the vocoder. RMSSinger is originally proposed for word-level realistic music score inputs, denoted as $\mathcal{S}$. To suit our settings, we drop the word-level attention module and directly use the fine-grained MIDI input. The alignment between MIDI notes and phonemes and other settings are reproduced according to \citet{he2023rmssinger}.

\section{Experiments}

In this section, we begin by showcasing experiments on AST tasks, followed by simulations and comparisons of a comprehensive annotation-and-training pipeline for an SVS task. We also investigated the model's performance in low-resource scenarios; however, due to space limitations, this part is included in \autoref{sec:low}.

\subsection{Experimental Setup}

% \subsubsection{Data}

\noindent \textbf{Data} \ We utilize two Mandarin datasets. The first is M4Singer \cite{zhang2022m4singer}, a multi-singer and multi-style singing voice corpus, which is approximately 26.5 hours after pre-processing. Secondly, we collect and annotate a high-quality song corpus, denoted as $\mathcal{D}_1$. $\mathcal{D}_1$ is composed of songs sung by 12 professional singers, with a total length of 20.9 hours. 
% These two datasets have the same structure, with $\mathcal{D}_1$ being considered an extension of M4Singer. 
For training AST models, these two datasets are used jointly, with two 3\% subsets used as the validation and the testing sets. Two 1\% subsets of MUSAN noise corpus are also isolated. The details of data collection are listed in \autoref{sec:data}.

% \subsubsection{Implementation and Training}
\vspace{2pt}

\noindent \textbf{Implementation and Training} \ The sample rate of waveforms is 24 kHz. F0 contours are extracted through a pre-trained RMVPE \cite{wei2023rmvpe} estimator, where each F0 value is quantized into 256 categories. The length of softened boundaries is set to 80 ms. The U-Net backbone is constructed with 4 down- and up-sampling layers, with $16 \times$ downsampling rate. For inference, the boundary threshold $\mu$ is set to 0.8. 
% More details are listed in \autoref{sec:arch}.
% We train the model for 60k steps using 2 NVIDIA 2080Ti GPUs with a batch size of 60k max frames. An AdamW optimizer is used with $\beta_1 = 0.9$, $\beta_2 = 0.98, \epsilon = 10^{-8}$. The learning rate is set to $10^{-5}$ with a decay rate of 0.998 and a decay step of 500 steps. 
More details are listed in \autoref{sec:arch}.

% \subsubsection{Evaluation}
\vspace{2pt}

\noindent \textbf{Evaluation} \ We utilize the \texttt{mir\_eval} library \cite{raffel2014mir_eval} for performance evaluation. 
% Specifically, we compute F1, precision, and recall scores of onset, offset, and pitch value. 
Specifically, we apply the metrics proposed in \cite{molina2014evaluation}: COn (correct onset), COff (correct offset), COnP (correct onset and pitch), COnPOff (correct onset, pitch, and offset).
An average overlap ratio (AOR) is also calculated for correctly transcribed note duration, and a raw pitch accuracy (RPA) for overall perception performance. 
% Specific implementations are discussed in \autoref{sec:ast-eval}.
% In addition, we compute the melody measures to reflect the overall perception performance, by transforming the GT and predicted note events into frame-level and computing raw pitch accuracy (RPA). For ROSVOT, we remove silence notes and designate the boundaries that enclose each note as onset and offset. This step is unnecessary for other baselines. The onset tolerance is set to 50 ms, and the offset tolerance is the larger value between 50 ms and 20\% of note duration. The pitch tolerance is set to 50 cents. All numbers demonstrated are multiplied by 100.
Given GT and corresponding predicted notes, their overlap ratio (OR) is defined as the ratio between the duration of the time segment in which the two notes overlap and the time segment spanned by the two notes combined. The AOR is given by the mean OR computed over all matching GT and predicted notes. 
For RPA, we transform the GT and predicted note events into frame-level sequences and compute matching scores. 

We remove silence notes and designate the boundaries that enclose each note as onset and offset for the evaluation of ROSVOT. This step is unnecessary for other baselines. The onset tolerance is set to 50 ms, and the offset tolerance is the larger value between 50 ms and 20\% of note duration. The pitch tolerance is set to 50 cents. All numbers demonstrated are multiplied by 100.

% \subsubsection{Baselines}
\vspace{2pt}

\noindent \textbf{Baselines} \ We compare ROSVOT, denoted as $\mathcal{M}$, with multiple baselines: 1) \textit{TONY} \cite{mauch2015computer}, an automatic software with visualization; 2) \textit{VOCANO} \cite{hsu2021vocano}, retrained on the joint datasets; 3) \textit{MusicYOLO}, retrained; 4) \textit{\cite{yong2023phoneme}}, reproduced and retrained. We also compare the results of several variants of $\mathcal{M}$: 
1) \textit{$\mathcal{M}$ (conformer)}, where the U-Net is dropped and the backbone is the Conformer alone; 
2) \textit{$\mathcal{M}$ (conv)}, where the middle Conformer blocks are replaced by 8-layer convolution blocks; 
3) \textit{$\mathcal{M}$ (w/o wbd)}, canceling word boundary condition;
4) \textit{$\mathcal{M}$ (w/o noise)}, which is identical to $\mathcal{M}$ but trained without noisy environment; 
5) \textit{$\mathcal{M}$ (w/ $E_\text{W}$)}, meaning that the GT word boundaries are not available and need to be extracted from the extractor $\text{E}_{\text{W}}$. 

\subsection{Main Results}

\begin{table*}
\setlength\tabcolsep{10pt}
\centering
\setlength{\belowcaptionskip}{-0.4cm}
\setlength\tabcolsep{4pt}
\begin{tabular*}{0.98\hsize}{l|cc|cc|cc|cc|cc}
\toprule
\multirow{2}{*}{ \bf Method }   & \multicolumn{2}{c|}{ \bf COn (F) $\uparrow$ }  & \multicolumn{2}{c|}{ \bf COff (F) $\uparrow$ } & \multicolumn{2}{c|}{ \bf COnPOff (F) $\uparrow$ } & \multicolumn{2}{c|}{ \bf Pitch (AOR) $\uparrow$ }  & \multicolumn{2}{c}{ \bf Melody (RPA) $\uparrow$ }        \\
                                & clean         & noisy             & clean         & noisy             & clean     & noisy     & clean     & noisy      & clean        & noisy   \\
\midrule
\textit{TONY}                   & 67.5          & 49.2              & 57.8          & 47.0              & 43.9      & 28.4      & 73.8      & 46.6        & 73.9        & 45.2  \\
\textit{VOCANO}                 & 75.8          & 64.7              & 71.2          & 66.1              & 50.2      & 43.4      & 81.4      & 71.9        & 76.6        & 59.8   \\
\textit{MusicYOLO}              & 82.2          & 79.7              & 81.7          & 76.5              & 58.9      & 51.5      & 85.4      & 78.6        & 81.6        & 78.9    \\
\textit{\cite{yong2023phoneme}} & 92.0          & 88.5              & 91.4          & 89.7              & 65.8      & 62.1      & 91.6      & 86.4        & 83.1        & 80.6    \\
\midrule[0.2pt]
\textit{$\mathcal{M}$ (conformer)}   & 92.1          & 90.6              & 91.8          & 90.8              & 70.3      & 69.8      & 95.9       & 95.3       & 83.9        & 83.1          \\
\textit{$\mathcal{M}$ (conv)}   & 91.6          & 91.5              & 92.6          & 92.6              & 70.9      & 70.8      & 96.8       & 96.8       & 84.1        & 84.1          \\
\textit{$\mathcal{M}$ (w/o wbd)}& 91.3          & 91.1              & 91.8          & 91.2              & 70.2      & 69.9      & 95.5       & 95.1       & 83.8        & 83.4          \\
\textit{$\mathcal{M}$ (w/o noise)}   & 93.8          & 90.9              & 94.2          & 91.5              & 76.4      & 70.1      & \bf97.1       & 95.2       & 87.1        & 83.1          \\
\textit{$\mathcal{M}$ (w/ $E_\text{W}$)}& 93.3          & 93.5              & 93.2          & 92.9              & 77.1      & 77.0      & 96.5       & 96.2       & 87.5        & 87.2          \\
\midrule[0.2pt]
\textit{$\mathcal{M}$ (ours)}   & \bf94.0          & \bf93.8              & \bf94.5          & \bf94.4              & \bf77.4      & \bf77.0      & 97.0       & \bf97.1       & \bf87.6        & \bf87.4          \\
\bottomrule
\end{tabular*}
\caption{\label{tab:main-results}
Evaluation results of AST systems. 
}
\end{table*}

We run two sets of experiments under clean and noisy environments, respectively. The noisy environment is produced by mixing MUSAN noises with a probability of 0.8 and an SNR range of $[6, 20]$. The main results are listed in \autoref{tab:main-results}. For the sake of brevity, only major scores are listed here, and the complete scores are listed in \autoref{sec:add}.

From the results, we can see that 1) the proposed multi-scale model achieves better performances for both boundary detection and pitch prediction by a large margin, even without noises; 2) Training under a noisy environment significantly improves the robustness, while the performances of baselines are severely degraded when facing noisy inputs; 3) The involvement of noises in training stage also improves the inference performance facing clean waveforms, this may because the noise mixing operation forms a bottleneck to force the model to focus on note-related information. 

\subsection{Ablation Study}

To demonstrate the effectiveness of several designs in the proposed method, we conduct ablation studies and compare the results of different hyperparameters. From \autoref{tab:main-results} we can see that dropping the U-Net backbone or replacing the Conformer with convolution blocks decreases the performance. In particular, the performance of \textit{$\mathcal{M}$ (conformer)} significantly deteriorates when dealing with noisy inputs, suggesting that the downsampling layers contribute to a denoising effect. This is also validated in the results of \textit{$\mathcal{M}$ (w/o noise)}, indicating that even though it is trained with clean samples, it still exhibits a certain level of robustness. For a fair comparison, we test \textit{$\mathcal{M}$ (w/ $E_\text{W}$)} to demonstrate the performance in note-only scenarios. The results indicate that despite the accumulated errors introduced by the word boundary extractor $\text{E}_{\text{W}}$, the performance does not decline significantly.

\begin{table}[ht]
\centering
\setlength{\belowcaptionskip}{-0.6cm}
\begin{tabular*}{0.98\hsize}{cc|ccc}
% \begin{tabular*}{0.8\hsize}{@{}@{\extracolsep{\fill}}l|cc|c@{}}
\toprule
\bf Rate          & \bf Step (ms)  & \bf COn       & \bf COff        & \bf COnPOff  \\
\midrule
% 1                  & 5.3            & -             & -               & -   \\
2                  & 10.7           & 92.7          & 92.4            & 70.7   \\
4                  & 21.3           & 93.9          & 93.6            & 73.6   \\
8                  & 42.7           & \bf94.4       & 94.1            & 76.8   \\
16                 & 85.3           & 94.0          & \bf94.5         & \bf77.4   \\
32                 & 170.7          & 94.3          & 94.1            & 77.2   \\
\bottomrule
\end{tabular*}
\caption{\label{tab:downsample}
Comparisons of different downsampling rates. "Step" denotes the downsampled step size in the Conformer, measured in milliseconds.
}
\end{table}

We record the comparison results of different overall downsampling rates of the U-Net backbone in \autoref{tab:downsample}, where only F1 scores are listed. The results align remarkably well with the length of the soft labels, which are both about 80 ms. We choose the rate of 16 in the final architecture $\mathcal{M}$ as it achieves better overall performance. 

For pitch prediction, we compare the results between the proposed attention-based method and the weighted median method used in previous works. We drop the pitch decoder $\text{D}_{\text{P}}$ and apply a weighted median algorithm on the F0 contours according to \cite{yong2023phoneme}. The F1 and AOR scores of this algorithm with clean inputs are 70.5 and 92.6, while the scores with noisy inputs are 63.2 and 86.6. The results indicate that a simple weighted median is insufficient in dealing with fluctuated pitches in singing voices, which are full of expressive techniques like portamentos. Also, its performance is largely dependent on the F0 extractor. 
% The attention-based decoder is designed to adjust the weights of features dynamically, capturing pitch information more effectively and robustly. 

\subsection{Towards Automatic Annotation}

The experimental results indicate that ROSVOT achieves superior performance, but what practical significance does it hold? In this section, we establish a comprehensive SVS pipeline, using ROSVOT as the automatic annotator. 

\begin{table*}
\setlength\tabcolsep{10pt}
\centering
\setlength{\belowcaptionskip}{-0.4cm}
\setlength\tabcolsep{3.5pt}
\begin{tabular*}{\hsize}{cccc|cc|cc|cc}
\toprule
\multirow{2}{*}{\bf Real}    & \multirow{2}{*}{\bf R. Size}  & \multirow{2}{*}{\bf Pseudo}       & \multirow{2}{*}{\bf P. Size}   &  \multicolumn{2}{c|}{\bf RPA}  & \multicolumn{2}{c|}{\bf MOS-P}  & \multicolumn{2}{c}{\bf MOS-Q}  \\
           &            &           &          & R & P & R & P & R & P     \\
\midrule
-       & -      & -     & -     & \multicolumn{2}{c|}{ 95.5 } & \multicolumn{2}{c|}{ 4.16$\pm$0.11 } & \multicolumn{2}{c}{ 4.08$\pm$0.08 }    \\
\midrule[0.2pt]
$\mathcal{D}_1$            & 20.9   & -                          & 0.0   & 67.6  & 61.1  & \bf3.69$\pm$0.09 & 3.54$\pm$0.07 & \bf3.81$\pm$0.04  & 3.74$\pm$0.05     \\
$\mathcal{D}_1\times50\%$  & 10.5   & $\mathcal{D}_1\times50\%$  & 10.5  & 66.0  & 61.0  & 3.62$\pm$0.03 & 3.56$\pm$0.08 & 3.76$\pm$0.07  & 3.72$\pm$0.03    \\
$\mathcal{D}_1\times10\%$  & 2.1    & $\mathcal{D}_1\times90\%$  & 18.8  & 65.9  & 61.3  & 3.65$\pm$0.05 & 3.55$\pm$0.05 & 3.71$\pm$0.08  & 3.73$\pm$0.04     \\
$\mathcal{D}_1\times5\%$   & 1.0    & $\mathcal{D}_1\times95\%$  & 19.9  & 63.0  & 63.3  & 3.61$\pm$0.05 & 3.57$\pm$0.04 & 3.73$\pm$0.05  & \bf3.78$\pm$0.03      \\
$\mathcal{D}_1\times1\%$   & 0.2    & $\mathcal{D}_1\times99\%$  & 20.7  & 63.5  & 64.7  & 3.60$\pm$0.04 & \bf3.59$\pm$0.04 & 3.74$\pm$0.06  & 3.76$\pm$0.04     \\
  -                        & 0.0    & $\mathcal{D}_1$            & 20.9  & 61.8  & 64.6  & 3.60$\pm$0.06 & 3.58$\pm$0.03 & 3.73$\pm$0.04  & 3.73$\pm$0.08     \\
\midrule[0.2pt]
M4                   & 26.5 & -                    & 0.0    & 68.1  & 67.7  & 3.63$\pm$0.05 & 3.60$\pm$0.04 & 3.79$\pm$0.05 & 3.77$\pm$0.07    \\
M4 + $\mathcal{D}_1$ & 47.4 & -                    & 0.0    & 67.4  & 66.5  & \bf3.67$\pm$0.07 & 3.59$\pm$0.08 & 3.81$\pm$0.04 & 3.80$\pm$0.06    \\
M4                   & 26.5 & $\mathcal{D}_1$      & 20.9   & 66.6  & 64.9  & 3.64$\pm$0.08 & \bf3.61$\pm$0.04 & 3.80$\pm$0.07 & 3.80$\pm$0.04    \\
M4                   & 26.5 & $\mathcal{D}_1$ + OP & 105.7  & 66.1  & 64.1  & 3.63$\pm$0.10 & 3.60$\pm$0.07 & \bf3.83$\pm$0.09 & \bf3.81$\pm$0.08    \\
% \midrule[0.2pt]
% M4        & M4          & 26.5 ($19\%$)          & $\mathcal{D}_1$ + OP + $\mathcal{D}_2$  & 111.7              & -        & -      & -       \\
% M4+$\mathcal{D}_1$        & M4+$\mathcal{D}_1$          & 47.4 ($34\%$)          & OP+$\mathcal{D}_2$  & 90.8              & -        & -      & -       \\
\bottomrule
\end{tabular*}
\caption{\label{tab:svs}
Evaluation results of SVS pipelines. The first row is for \textit{GT Mel}, where we generate waveforms using the vocoder from GT Mel-spectrograms. "M4" denotes M4Singer, and "OP" denotes OpenSinger. "R" and "P" denote inference results using real or pseudo annotations, respectively. The sizes are measured in hours. 
}
\end{table*}

\begin{table}[ht]
\centering
\setlength{\belowcaptionskip}{-0.6cm}
\begin{tabular*}{0.97\hsize}{c|ccc}
\toprule
\bf Model                       & \bf RPA          & \bf MOS-P        & \bf MOS-Q     \\
\midrule
\textit{GT}                     & 96.1             & 4.02 $\pm$ 0.05    & 4.04 $\pm$ 0.06  \\
\textit{$\mathcal{S}$(large)}   & 45.2             & 3.36 $\pm$ 0.12    & 3.45 $\pm$ 0.09  \\
\bottomrule
\end{tabular*}
\caption{\label{tab:cross}
Results of cross-lingual SVS generalization.
}
\end{table}

\subsubsection{Implementation and Pipeline}

\noindent \textbf{Data.} We re-align and re-annotate the OpenSinger corpus \cite{huang2021multi}, which consists of 84.8 hours of singing voices recorded by 93 singers. We also perform cross-lingual generalization by annotating an English corpus $\mathcal{D}_2$, which has a length of 6 hours.
% We recruit 8 professional singers to record $\mathcal{D}_2$, and then automatically annotate the phonemes and notes through an ASR model \cite{radford2023robust}, MFA, and the proposed AST model. The length of $\mathcal{D}_2$ is 6 hours. 
For future reference, we use the term pseudo-annotations for the automatically generated transcriptions. Details are listed in \autoref{sec:data}.

% \noindent \textbf{SVS model.} We choose RMSSinger as the singing acoustic model and a pre-trained HiFi-GAN \cite{kong2020hifi} model as the vocoder. RMSSinger is originally proposed for word-level realistic music score inputs, denoted as $\mathcal{S}$. To suit our settings, we drop the word-level attention module and directly use the fine-grained MIDI input. The alignment between MIDI notes and phonemes and other settings are reproduced according to \citet{he2023rmssinger}.

\noindent \textbf{Evaluation.} For objective evaluation, we also apply the RPA score to measure the reconstructed F0 contours. The RPA scores for \textit{GT Mel} are computed between F0s from GT vocoder generations and GT waveforms, while the others are between GT and generations. For subjective evaluation, we conducted crowdsourced mean opinion score (MOS) listening tests. Specifically, we score MOS-P and MOS-Q corresponding to pitch reconstruction and overall quality. The metrics are rated from 1 to 5 and reported with 95\% confidence intervals. For a more intuitive demonstration, we record Comparative Mean Opinion Scores (CMOS) and discuss the results in \autoref{sec:add}.

\subsubsection{SVS Results}

Firstly, we investigate the effect of training with pseudo-annotations at different ratios.
% on the final synthesis performance. 
We only utilize M4Singer to train ROSVOT $\mathcal{M}$, which is used to generate the pseudo annotations. Pseudo annotations with different ratios are mixed into $\mathcal{D}_1$ to form the training set. For inference, we reserve two 1\% segments from each real and pseudo group for validation and testing. The results are listed in \autoref{tab:svs}, rows 2-7. From the results, we can see that the pitch accuracy of real annotation inputs decreases when mixing more pseudo annotations, but the accuracy of pseudo inputs increases. This suggests a minor discrepancy in the distributions of real and pseudo annotations. However, the performance degradation is not significant: 99\% of the pseudo mixing contributes only a 6\% drop in performance. The MOS-Q scores share a similar pattern, but they involve a comprehensive evaluation with considerations of audio quality and more. A decrease in pitch accuracy does not necessarily lead to an overall decline in quality. 

We further investigate the performance as data size increases. While the AST model remains the same, we train $\mathcal{S}$ only using M4Singer as the baseline. Next, we gradually mix $\mathcal{D}_1$ and OpenSinger to expand the data size. To consume the largest datasets in the last row, we construct a large version of RMSSinger with 320-dimensional channels and a 6-layer decoder\footnote{The dictionary of the text encoder is also merged with English phonemes for the following cross-lingual experiments.}, denoted as \textit{$\mathcal{S}$(large)}. The results are listed in \autoref{tab:svs}, rows 8-11. A slight reduction in pitch accuracy can be observed when integrating diverse datasets, which may result from the inherent differences in dataset characteristics and annotation styles. However, the overall quality slightly improves, as the model has been exposed to a sufficient variety of pronunciation styles and singing patterns. This indicates that ROSVOT provides an opportunity for SVS models to scale up, which could be beneficial for large-scale singing voice pre-training or zero-shot SVS.

\subsubsection{Cross-lingual Generalization}

We further investigate the zero-shot cross-lingual capability of ROSVOT, and explore the feasibility of generalizing a Mandarin SVS model to English. We use the same ROSVOT model combined with $\text{E}_{\text{W}}$ trained with M4Singer and test it on TONAS \cite{gomez2013towards}, a flamenco a cappella singing corpus. The quantitative results are listed in \autoref{tab:comp} in \autoref{sec:add}. Performance is degraded, and we believe it may be due to the flamenco singing style (which has rich techniques like appoggiatura) and the unseen language, since the model is trained with Mandarin pop songs. 

For SVS, we finetune the pre-trained model \textit{$\mathcal{S}$(large)} on the English corpus $\mathcal{D}_2$, a 6-hour dataset transcribed automatically using MFA and the AST model above. Note that since the word duration information is obtained using MFA, we drop $\text{E}_{\text{W}}$ and directly generate word-note synchronized transcriptions. We finetune both stages 1 and 2 of RMSSinger for 100k steps. The evaluation results are listed in \autoref{tab:cross}.

Although the performance is not superior, the results still demonstrate certain cross-lingual capabilities. There are vast differences in pronunciation rules and phonetic characteristics between English and Mandarin, since they belong to two distinct language families. If a model possesses a certain capability to transfer from Mandarin to English, it essentially demonstrates a degree of general cross-linguistic ability. We will leave elaborate validations and evaluations of general cross-lingual capabilities for future work.

\section{Conclusion}

In this paper, we introduced ROSVOT, the first robust AST model that ultimately serves SVS. We leveraged a multi-scale architecture to achieve a balance between coarse-grained note modeling and fine-grained segmentation. An attention-based decoder with dynamic weight was devised for pitch regression. Additionally, we established a comprehensive pipeline for SVS training. Experimental results revealed that our model achieved the best performance under either clean or noisy environments. Annotating and incorporating larger datasets improved the SVS model's performance, indicating the capability of practical annotation of ROSVOT. 

\section*{Limitations and Potential Risks}

% The proposed method primarily acknowledges two key limitations. Firstly, we only test the cross-lingual capability of the proposed model on a small-scale English dataset. Extensional experiments are required to fully measure the generalization performance. Secondly, due to the space limitation, only one SVS model is tested as the baseline. More verifications of different SVS models are needed to demonstrate the practical performance. In the future, we plan to test the automatically annotated transcriptions on more diverse SVS models. 
The proposed method acknowledges two primary limitations. First, the cross-lingual capability is only tested on a small-scale English dataset, necessitating extensional experiments for a comprehensive evaluation of generalization performance. Second, due to space constraints, only one SVS model is examined as the baseline. Additional verifications involving different SVS models are required to fully demonstrate practical performance. Future work will involve testing automatically annotated transcriptions on a more diverse set of SVS models.

The misuse of the proposed model for singing voice synthesis could potentially lead to copyright-related issues. To address this concern, appropriate constraints will be implemented to mitigate any illegal or unauthorized usage.

\section*{Acknowledgement}

This work was supported in part by the National Natural Science Foundation of China under Grant No.62222211.

% Bibliography entries for the entire Anthology, followed by custom entries
%\bibliography{anthology,custom}
% Custom bibliography entries only
\bibliography{acl_latex}

\begin{thebibliography}{38}
\expandafter\ifx\csname natexlab\endcsname\relax\def\natexlab#1{#1}\fi

\bibitem[{Baevski et~al.(2020)Baevski, Zhou, Mohamed, and Auli}]{baevski2020wav2vec}
Alexei Baevski, Yuhao Zhou, Abdelrahman Mohamed, and Michael Auli. 2020.
\newblock wav2vec 2.0: A framework for self-supervised learning of speech representations.
\newblock \emph{Advances in neural information processing systems}, 33:12449--12460.

\bibitem[{Bhattarai and Lee(2023)}]{bhattarai2023comprehensive}
Bhuwan Bhattarai and Joonwhoan Lee. 2023.
\newblock A comprehensive review on music transcription.
\newblock \emph{Applied Sciences}, 13(21):11882.

\bibitem[{Chen et~al.(2020)Chen, Tan, Luan, Qin, and Liu}]{chen2020hifisinger}
Jiawei Chen, Xu~Tan, Jian Luan, Tao Qin, and Tie-Yan Liu. 2020.
\newblock Hifisinger: Towards high-fidelity neural singing voice synthesis.
\newblock \emph{arXiv preprint arXiv:2009.01776}.

\bibitem[{Conneau et~al.(2020)Conneau, Baevski, Collobert, Mohamed, and Auli}]{conneau2020unsupervised}
Alexis Conneau, Alexei Baevski, Ronan Collobert, Abdelrahman Mohamed, and Michael Auli. 2020.
\newblock Unsupervised cross-lingual representation learning for speech recognition.
\newblock \emph{arXiv preprint arXiv:2006.13979}.

\bibitem[{Dai et~al.(2019)Dai, Yang, Yang, Carbonell, Le, and Salakhutdinov}]{dai2019transformer}
Zihang Dai, Zhilin Yang, Yiming Yang, Jaime Carbonell, Quoc~V Le, and Ruslan Salakhutdinov. 2019.
\newblock Transformer-xl: Attentive language models beyond a fixed-length context.
\newblock \emph{arXiv preprint arXiv:1901.02860}.

\bibitem[{Fu and Su(2019)}]{fu2019hierarchical}
Zih-Sing Fu and Li~Su. 2019.
\newblock Hierarchical classification networks for singing voice segmentation and transcription.
\newblock In \emph{Proceedings of the 20th International Society for Music Information Retrieval Conference (ISMIR 2019)}, pages 900--907.

\bibitem[{G{\'o}mez and Bonada(2013)}]{gomez2013towards}
Emilia G{\'o}mez and Jordi Bonada. 2013.
\newblock Towards computer-assisted flamenco transcription: An experimental comparison of automatic transcription algorithms as applied to a cappella singing.
\newblock \emph{Computer Music Journal}, 37(2):73--90.

\bibitem[{Gu et~al.(2023)Gu, Zeng, Zhang, Ou, and Wang}]{gu2023deep}
Xiangming Gu, Wei Zeng, Jianan Zhang, Longshen Ou, and Ye~Wang. 2023.
\newblock Deep audio-visual singing voice transcription based on self-supervised learning models.
\newblock \emph{arXiv preprint arXiv:2304.12082}.

\bibitem[{Gulati et~al.(2020)Gulati, Qin, Chiu, Parmar, Zhang, Yu, Han, Wang, Zhang, Wu et~al.}]{gulati2020conformer}
Anmol Gulati, James Qin, Chung-Cheng Chiu, Niki Parmar, Yu~Zhang, Jiahui Yu, Wei Han, Shibo Wang, Zhengdong Zhang, Yonghui Wu, et~al. 2020.
\newblock Conformer: Convolution-augmented transformer for speech recognition.
\newblock \emph{arXiv preprint arXiv:2005.08100}.

\bibitem[{He et~al.(2023)He, Liu, Ye, Huang, Cui, Liu, and Zhao}]{he2023rmssinger}
Jinzheng He, Jinglin Liu, Zhenhui Ye, Rongjie Huang, Chenye Cui, Huadai Liu, and Zhou Zhao. 2023.
\newblock Rmssinger: Realistic-music-score based singing voice synthesis.
\newblock \emph{arXiv preprint arXiv:2305.10686}.

\bibitem[{Hsu et~al.(2021)Hsu, Su et~al.}]{hsu2021vocano}
Jui-Yang Hsu, Li~Su, et~al. 2021.
\newblock Vocano: A note transcription framework for singing voice in polyphonic music.

\bibitem[{Huang et~al.(2021)Huang, Chen, Ren, Liu, Cui, and Zhao}]{huang2021multi}
Rongjie Huang, Feiyang Chen, Yi~Ren, Jinglin Liu, Chenye Cui, and Zhou Zhao. 2021.
\newblock Multi-singer: Fast multi-singer singing voice vocoder with a large-scale corpus.
\newblock In \emph{Proceedings of the 29th ACM International Conference on Multimedia}, pages 3945--3954.

\bibitem[{Huang et~al.(2022)Huang, Cui, Chen, Ren, Liu, Zhao, Huai, and Wang}]{huang2022singgan}
Rongjie Huang, Chenye Cui, Feiyang Chen, Yi~Ren, Jinglin Liu, Zhou Zhao, Baoxing Huai, and Zhefeng Wang. 2022.
\newblock Singgan: Generative adversarial network for high-fidelity singing voice generation.
\newblock In \emph{Proceedings of the 30th ACM International Conference on Multimedia}, pages 2525--2535.

\bibitem[{Kim et~al.(2021)Kim, Kong, and Son}]{kim2021vits}
Jaehyeon Kim, Jungil Kong, and Juhee Son. 2021.
\newblock Vits: Conditional variational autoencoder with adversarial learning for end-to-end text-tospeech.
\newblock In \emph{Proc. ICML}, pages 5530--5540.

\bibitem[{Kong et~al.(2020)Kong, Kim, and Bae}]{kong2020hifi}
Jungil Kong, Jaehyeon Kim, and Jaekyoung Bae. 2020.
\newblock Hifi-gan: Generative adversarial networks for efficient and high fidelity speech synthesis.
\newblock \emph{Advances in Neural Information Processing Systems}, 33:17022--17033.

\bibitem[{Li et~al.(2023)Li, Huang, Zhang, Liu, and Zhao}]{li2023alignsts}
Ruiqi Li, Rongjie Huang, Lichao Zhang, Jinglin Liu, and Zhou Zhao. 2023.
\newblock Alignsts: Speech-to-singing conversion via cross-modal alignment.
\newblock \emph{arXiv preprint arXiv:2305.04476}.

\bibitem[{Lin et~al.(2017)Lin, Goyal, Girshick, He, and Doll{\'a}r}]{lin2017focal}
Tsung-Yi Lin, Priya Goyal, Ross Girshick, Kaiming He, and Piotr Doll{\'a}r. 2017.
\newblock Focal loss for dense object detection.
\newblock In \emph{Proceedings of the IEEE international conference on computer vision}, pages 2980--2988.

\bibitem[{Liu et~al.(2022)Liu, Li, Ren, Chen, and Zhao}]{Liu_Li_Ren_Chen_Zhao_2022}
Jinglin Liu, Chengxi Li, Yi~Ren, Feiyang Chen, and Zhou Zhao. 2022.
\newblock \href {https://doi.org/10.1609/aaai.v36i10.21350} {Diffsinger: Singing voice synthesis via shallow diffusion mechanism}.
\newblock \emph{Proceedings of the AAAI Conference on Artificial Intelligence}, 36(10):11020--11028.

\bibitem[{Mauch et~al.(2015)Mauch, Cannam, Bittner, Fazekas, Salamon, Dai, Bello, and Dixon}]{mauch2015computer}
Matthias Mauch, Chris Cannam, Rachel Bittner, George Fazekas, Justin Salamon, Jiajie Dai, Juan Bello, and Simon Dixon. 2015.
\newblock Computer-aided melody note transcription using the tony software: Accuracy and efficiency.

\bibitem[{McAuliffe et~al.(2017)McAuliffe, Socolof, Mihuc, Wagner, and Sonderegger}]{mcauliffe2017montreal}
Michael McAuliffe, Michaela Socolof, Sarah Mihuc, Michael Wagner, and Morgan Sonderegger. 2017.
\newblock Montreal forced aligner: Trainable text-speech alignment using kaldi.
\newblock In \emph{Interspeech}, volume 2017, pages 498--502.

\bibitem[{Molina et~al.(2014)Molina, Barbancho-Perez, Tardon-Garcia, Barbancho-Perez et~al.}]{molina2014evaluation}
Emilio Molina, Ana~Maria Barbancho-Perez, Lorenzo~Jose Tardon-Garcia, Isabel Barbancho-Perez, et~al. 2014.
\newblock Evaluation framework for automatic singing transcription.

\bibitem[{Radford et~al.(2023)Radford, Kim, Xu, Brockman, McLeavey, and Sutskever}]{radford2023robust}
Alec Radford, Jong~Wook Kim, Tao Xu, Greg Brockman, Christine McLeavey, and Ilya Sutskever. 2023.
\newblock Robust speech recognition via large-scale weak supervision.
\newblock In \emph{International Conference on Machine Learning}, pages 28492--28518. PMLR.

\bibitem[{Raffel et~al.(2014)Raffel, McFee, Humphrey, Salamon, Nieto, Liang, Ellis, and Raffel}]{raffel2014mir_eval}
Colin Raffel, Brian McFee, Eric~J Humphrey, Justin Salamon, Oriol Nieto, Dawen Liang, Daniel~PW Ellis, and C~Colin Raffel. 2014.
\newblock mir\_eval: A transparent implementation of common mir metrics.
\newblock In \emph{In Proceedings of the 15th International Society for Music Information Retrieval Conference, ISMIR}. Citeseer.

\bibitem[{Ren et~al.(2020)Ren, Tan, Qin, Luan, Zhao, and Liu}]{ren2020deepsinger}
Yi~Ren, Xu~Tan, Tao Qin, Jian Luan, Zhou Zhao, and Tie-Yan Liu. 2020.
\newblock Deepsinger: Singing voice synthesis with data mined from the web.
\newblock In \emph{Proceedings of the 26th ACM SIGKDD International Conference on Knowledge Discovery \& Data Mining}, pages 1979--1989.

\bibitem[{Ronneberger et~al.(2015)Ronneberger, Fischer, and Brox}]{ronneberger2015u}
Olaf Ronneberger, Philipp Fischer, and Thomas Brox. 2015.
\newblock U-net: Convolutional networks for biomedical image segmentation.
\newblock In \emph{Medical Image Computing and Computer-Assisted Intervention--MICCAI 2015: 18th International Conference, Munich, Germany, October 5-9, 2015, Proceedings, Part III 18}, pages 234--241. Springer.

\bibitem[{Shen et~al.(2023)Shen, Ju, Tan, Liu, Leng, He, Qin, Zhao, and Bian}]{shen2023naturalspeech}
Kai Shen, Zeqian Ju, Xu~Tan, Yanqing Liu, Yichong Leng, Lei He, Tao Qin, Sheng Zhao, and Jiang Bian. 2023.
\newblock Naturalspeech 2: Latent diffusion models are natural and zero-shot speech and singing synthesizers.
\newblock \emph{arXiv preprint arXiv:2304.09116}.

\bibitem[{Singla et~al.(2022)Singla, Shah, Chen, and Shah}]{singla2022audio}
Yaman~Kumar Singla, Jui Shah, Changyou Chen, and Rajiv~Ratn Shah. 2022.
\newblock What do audio transformers hear? probing their representations for language delivery \& structure.
\newblock In \emph{2022 IEEE International Conference on Data Mining Workshops (ICDMW)}, pages 910--925. IEEE.

\bibitem[{Snyder et~al.(2015)Snyder, Chen, and Povey}]{musan2015}
David Snyder, Guoguo Chen, and Daniel Povey. 2015.
\newblock \href {http://arxiv.org/abs/1510.08484} {{MUSAN}: {A} {M}usic, {S}peech, and {N}oise {C}orpus}.
\newblock ArXiv:1510.08484v1.

\bibitem[{Wang and Jang(2021)}]{wang2021preparation}
Jun-You Wang and Jyh-Shing~Roger Jang. 2021.
\newblock On the preparation and validation of a large-scale dataset of singing transcription.
\newblock In \emph{ICASSP 2021-2021 IEEE International Conference on Acoustics, Speech and Signal Processing (ICASSP)}, pages 276--280. IEEE.

\bibitem[{Wang et~al.(2022{\natexlab{a}})Wang, Xu, Yang, and Cheng}]{wang2022musicyolo}
Xianke Wang, Wei Xu, Weiming Yang, and Wenqing Cheng. 2022{\natexlab{a}}.
\newblock Musicyolo: A sight-singing onset/offset detection framework based on object detection instead of spectrum frames.
\newblock In \emph{ICASSP 2022-2022 IEEE International Conference on Acoustics, Speech and Signal Processing (ICASSP)}, pages 396--400. IEEE.

\bibitem[{Wang et~al.(2022{\natexlab{b}})Wang, Wang, Zhu, Wu, Li, Xue, Zhang, Xie, and Bi}]{wang2022opencpop}
Yu~Wang, Xinsheng Wang, Pengcheng Zhu, Jie Wu, Hanzhao Li, Heyang Xue, Yongmao Zhang, Lei Xie, and Mengxiao Bi. 2022{\natexlab{b}}.
\newblock Opencpop: A high-quality open source chinese popular song corpus for singing voice synthesis.
\newblock \emph{arXiv preprint arXiv:2201.07429}.

\bibitem[{Wei et~al.(2023)Wei, Cao, Dan, and Chen}]{wei2023rmvpe}
Haojie Wei, Xueke Cao, Tangpeng Dan, and Yueguo Chen. 2023.
\newblock Rmvpe: A robust model for vocal pitch estimation in polyphonic music.
\newblock \emph{arXiv preprint arXiv:2306.15412}.

\bibitem[{Yang et~al.(2023)Yang, Tian, Tan, Huang, Liu, Chang, Shi, Zhao, Bian, Wu et~al.}]{yang2023uniaudio}
Dongchao Yang, Jinchuan Tian, Xu~Tan, Rongjie Huang, Songxiang Liu, Xuankai Chang, Jiatong Shi, Sheng Zhao, Jiang Bian, Xixin Wu, et~al. 2023.
\newblock Uniaudio: An audio foundation model toward universal audio generation.
\newblock \emph{arXiv preprint arXiv:2310.00704}.

\bibitem[{Yong et~al.(2023)Yong, Su, and Nam}]{yong2023phoneme}
Sangeon Yong, Li~Su, and Juhan Nam. 2023.
\newblock A phoneme-informed neural network model for note-level singing transcription.
\newblock In \emph{ICASSP 2023-2023 IEEE International Conference on Acoustics, Speech and Signal Processing (ICASSP)}, pages 1--5. IEEE.

\bibitem[{Zhang et~al.(2022{\natexlab{a}})Zhang, Li, Wang, Deng, Liu, Ren, He, Huang, Zhu, Chen et~al.}]{zhang2022m4singer}
Lichao Zhang, Ruiqi Li, Shoutong Wang, Liqun Deng, Jinglin Liu, Yi~Ren, Jinzheng He, Rongjie Huang, Jieming Zhu, Xiao Chen, et~al. 2022{\natexlab{a}}.
\newblock M4singer: A multi-style, multi-singer and musical score provided mandarin singing corpus.
\newblock \emph{Advances in Neural Information Processing Systems}, 35:6914--6926.

\bibitem[{Zhang et~al.(2022{\natexlab{b}})Zhang, Cong, Xue, Xie, Zhu, and Bi}]{9747664}
Yongmao Zhang, Jian Cong, Heyang Xue, Lei Xie, Pengcheng Zhu, and Mengxiao Bi. 2022{\natexlab{b}}.
\newblock \href {https://doi.org/10.1109/ICASSP43922.2022.9747664} {Visinger: Variational inference with adversarial learning for end-to-end singing voice synthesis}.
\newblock In \emph{ICASSP 2022 - 2022 IEEE International Conference on Acoustics, Speech and Signal Processing (ICASSP)}, pages 7237--7241.

\bibitem[{Zhang et~al.(2023)Zhang, Huang, Li, He, Xia, Chen, Duan, Huai, and Zhao}]{zhang2023stylesinger}
Yu~Zhang, Rongjie Huang, Ruiqi Li, JinZheng He, Yan Xia, Feiyang Chen, Xinyu Duan, Baoxing Huai, and Zhou Zhao. 2023.
\newblock Stylesinger: Style transfer for out-of-domain singing voice synthesis.
\newblock \emph{arXiv preprint arXiv:2312.10741}.

\bibitem[{Zhang et~al.(2022{\natexlab{c}})Zhang, Zheng, Li, and Lu}]{zhang2022wesinger}
Zewang Zhang, Yibin Zheng, Xinhui Li, and Li~Lu. 2022{\natexlab{c}}.
\newblock Wesinger: Data-augmented singing voice synthesis with auxiliary losses.
\newblock \emph{arXiv preprint arXiv:2203.10750}.

\end{thebibliography}

\appendix

\section{Word Boundary Condition} \label{sec:wbd}

Word boundary conditions are introduced to regulate segmentation results. It seems similar to \citet{yong2023phoneme}, but a word boundary sequence forms a much narrower information bottleneck without introducing unnecessary information. In practice, we embed the word boundary sequences to inform the model of boundary conditions. Also, we use the word boundary sequence as a reference to regulate the predicted note boundaries. Specifically, we remove the note boundaries that are too close to the reference word boundaries, where the threshold is 40 ms. 

This regulation is only for automatic annotation. For a note-only application, word-note synchronization is not necessary. In this scenario, we build a word boundary extractor $\text{E}_{\text{W}}$ to provide weak linguistic supervision. The extractor shares the same architecture as the note segmentation part of ROSVOT. The multi-scale architecture also functions well in localizing frame-level word boundaries. Specifically, we use an MFA-aligned AISHELL-3 Mandarin corpus to pre-train $\text{E}_{\text{W}}$, followed by fine-tuning it with M4Singer and $\mathcal{D}_1$.

\section{Architecture and Implementation Details} \label{sec:arch}

\subsection{Hyperparameters}

For hyperparameters, we sample waveforms with a sample rate of 24000 Hz. Mel-spectrograms are computed with a window size of 512, and a hop size of 128. The number of Mel bins is set to 80. To form a bottleneck and alleviate overfitting, we only use the first 30 bins (low-frequency part) as input. MUSAN noises are added to the waveforms with a probability of 0.8 and an SNR range of $[6, 20]$. Gaussian noise is added to the logarithmic F0 contours with a random standard deviation range of $[0, 0.04]$, and is added to the softened boundary labels with $[0, 0.002]$. F0 contours are extracted through a pre-trained RMVPE \cite{wei2023rmvpe} estimator, where each F0 value is quantized into 256 categories. We set $P$, the number of pitch categories, to 120, where each pitch number is the exact MIDI number. The length of softened boundaries is set to 80 ms, indicating a 15-frame window $W_{\mathcal{G}}$. The temperature parameters $T_1$ and $T_2$ are set to 0.2 and 0.01. To balance the various objectives, we set $\lambda_{\text{B}}$, $\lambda_{\text{FC}}$, and$\lambda_{\text{P}}$ to $1.0$, $3.0$, and $1.0$. For inference, the boundary threshold $\mu$ is set to 0.8. The hyperparameters $\alpha$ and $\gamma$ in the boundary decoder are set to $1/(2.42\times128/24000)$ and 5.0, where the $2.42$ in the former indicates the number of note boundaries in one second, and 128 and 24000 indicate the hop size and the audio sample rate. 

We train the AST model for 60k steps using 2 NVIDIA 2080Ti GPUs with a batch size of 60k max frames. An AdamW optimizer is used with $\beta_1 = 0.9$, $\beta_2 = 0.98, \epsilon = 10^{-8}$. The learning rate is set to $10^{-5}$ with a decay rate of 0.998 and a decay step of 500 steps. 

\subsection{Architecture}

For model architecture, we apply three encoders $\text{E}_{\text{M}}$, $\text{E}_{\text{B}}$, $\text{E}_{\text{P}}$ to encode Mel-spectrograms, word boundaries, and F0 contours. The encoders consist of a linear projection or an embedding layer, followed by residual convolution blocks. 

The U-Net backbone's encoder and decoder each comprise $K$ downsampling and upsampling layers, respectively, where $K=4$ in our case, with $16 \times$ downsampling rate. 
A downsampling layer consists of a residual convolution block and an average pooling layer with a downsampling rate of 2, resulting in an overall downsampling rate of $2^K$. For an upsampling layer, the input feature is firstly upsampled through a transposed convolution layer, and is then concatenated with the corresponding skipped feature before a final convolution block. The downsampling rate is set to 2, and the channel dimension remains the same as input to alleviate overfitting. 
The intermediate part of the backbone is replaced by a 2-layer Conformer block with relative position encoding.
% The U-Net backbone is constructed with 4 down- and up-sampling layers, with $16 \times$ downsampling rate. 
The Conformer network is 2-layer with a kernel size of 9 and a head size of 4. The head dimension in the pitch decoder is 4. The overall channel dimension is 256. The overall architecture is listed in \autoref{tab:arch}.

\begin{figure}[htbp]
    \centering
    \includegraphics[width=0.3\textwidth]{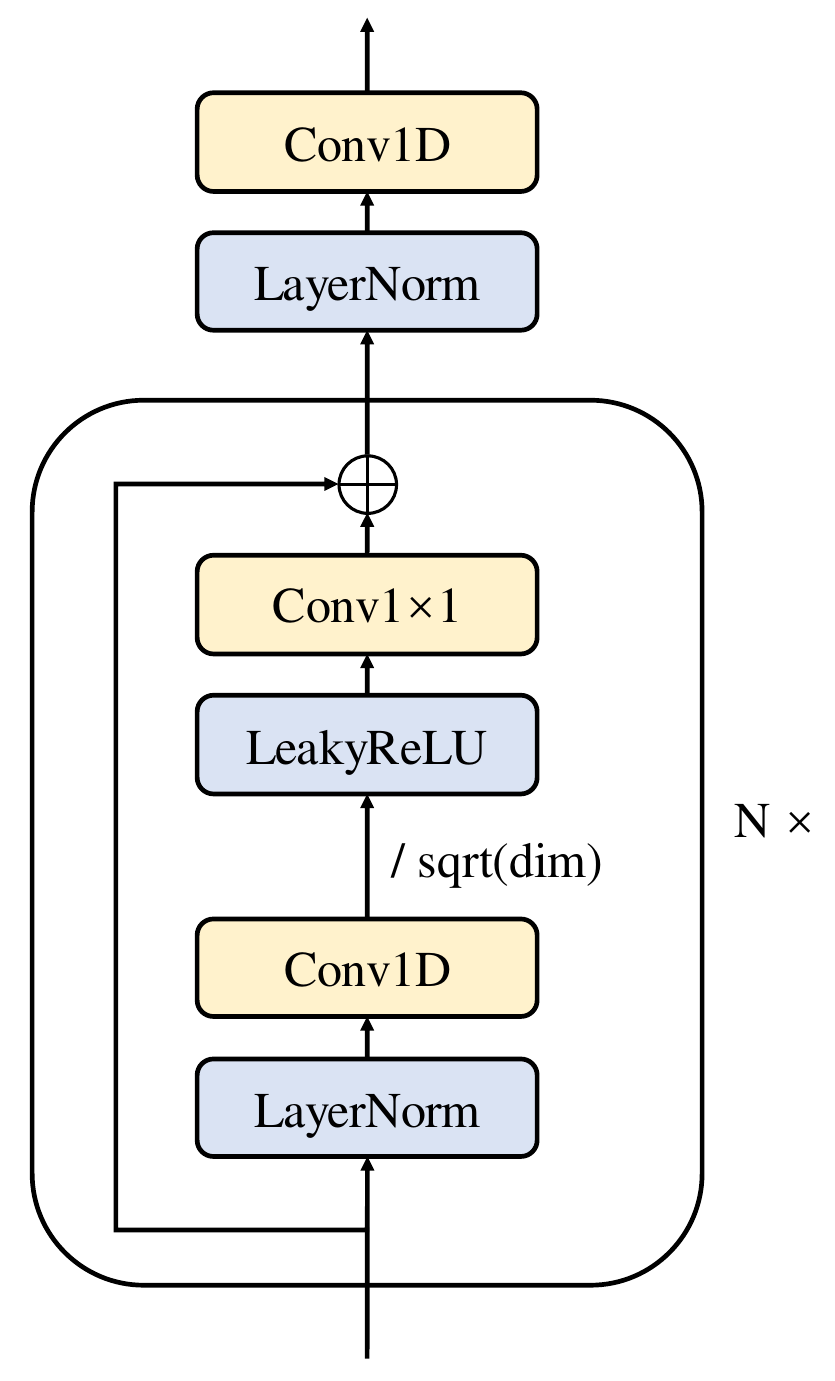}
    \setlength{\belowcaptionskip}{-0.5cm}
    \caption{
        N-layer Residual convolution blocks. 
    }
    \label{fig:conv}
\end{figure}

As for the N-layer residual convolution blocks mentioned many times in the main text, the configuration is illustrated in \autoref{fig:conv}. 

\begin{table}
\centering
\setlength{\belowcaptionskip}{-0.4cm}
\begin{tabular*}{\hsize}{l|c|c}
% \begin{tabular*}{0.8\hsize}{@{}@{\extracolsep{\fill}}l|cc|c@{}}
\toprule
\multicolumn{2}{c|}{Hyperparameter}                            & Model    \\
\midrule
\multirow{3}*{\shortstack{Mel\\Encoder}}                & Encoder Kernel            & 3     \\
~                                                       & Encoder Layers            & 2     \\
~                                                       & Encoder Hidden            & 256   \\
\midrule[0.2pt]
\multirow{6}*{\shortstack{Condition\\Encoder}}          & Pitch Embedding           & 300     \\
~                                                       & UV Embedding              & 3     \\
~                                                       & WBD Embedding             & 3     \\
~                                                       & Encoder Kernel            & 3     \\
~                                                       & Encoder Layers            & 1     \\
~                                                       & Encoder Hidden            & 256   \\
\midrule[0.2pt]
\multirow{4}*{\shortstack{U-Net}}                       & Kernel                    & 3     \\
~                                                       & Enc \& Dec Layers         & 4     \\
~                                                       & Downsampling Rate         & 16   \\
~                                                       & Enc \& Dec Hidden         & 256   \\
\midrule[0.2pt]
\multirow{5}*{\shortstack{Conformer}}                   & Kernel                    & 9     \\
~                                                       & Heads                     & 4     \\
~                                                       & Layers                    & 2   \\
~                                                       & Attention Hidden          & 256   \\
~                                                       & FFN Hidden                & 1024   \\
\midrule[0.2pt]
\multicolumn{2}{c|}{Total Number of Parameters}                     & 12M    \\
\bottomrule
\end{tabular*}
\caption{\label{tab:arch}
Hyperparameters of the proposed modules. "WBD" represents word boundary.
}
\end{table}

\section{Data} \label{sec:data}

We recruit 12 professional singers (8 female, 4 male) to record $\mathcal{D}_1$ and 8 singers (5 female, 3 male) for $\mathcal{D}_2$. Each singer was compensated at an hourly rate of \$600. Singers were informed that the recordings were for scientific research use. For $\mathcal{D}_1$, we have hired music experts to manually annotate the note and word information. Each annotator was compensated at an hourly rate of \$20. Participants were informed that the data would be used for scientific research. For $\mathcal{D}_2$, we automatically annotate the words and notes through an ASR model \cite{radford2023robust}, MFA, and the proposed AST model. The length of $\mathcal{D}_1$ is 20.9 hours and $\mathcal{D}_2$ is 6 hours. We use all the datasets under license CC BY-NC-SA 4.0.

\section{Objectives} \label{sec:loss}

The binary cross-entropy (BCE) loss applied to train the note segmentation stage:

\begin{align}
    \mathcal{L}_{\text{B}} = & \frac{1}{T} \sum \text{ BCE}(\widetilde{\boldsymbol{y}}_{\text{bd}}, \hat{\boldsymbol{y}}_{\text{bd}}) \nonumber \\
                  = & - \frac{1}{T} \sum_{t=1}^{T} ( \widetilde{y}_{\text{bd}}^t \ln(\sigma(\hat{y}_{\text{bd}}^t / T_1)) \nonumber \\
            & + (1 - \widetilde{y}_{\text{bd}}^t) \ln(1 - \sigma(\hat{y}_{\text{bd}}^t / T_1)) )
\end{align}
where $T_1$ is the temperature hyperparameter, and $\sigma(\cdot)$ stands for sigmoid function.

To tackle the imbalance problem, we employ a focal loss \cite{lin2017focal}, $\mathcal{L}_{\text{FC}}$, to focus more on hard samples:
\begin{align}
    & p_{\text{t}} = \boldsymbol{y}_{\text{bd}} \sigma(\hat{\boldsymbol{y}}_{\text{bd}}) + (1 - \boldsymbol{y}_{\text{bd}}) (1 - \sigma(\hat{\boldsymbol{y}}_{\text{bd}})) \\
    & \alpha_{\text{t}} = \alpha \boldsymbol{y}_{\text{bd}} + (1 - \alpha) (1 - \boldsymbol{y}_{\text{bd}}) \\
    & \mathcal{L}_{\text{FC}} = \frac{1}{T} \sum \alpha_{\text{t}} (1 - p_{\text{t}})^{\gamma} \text{BCE}(\boldsymbol{y}_{\text{bd}}, \hat{\boldsymbol{y}}_{\text{bd}})
\end{align}
where $\alpha$ is a hyperparameter controlling weight of positive samples, and $\gamma$ controls balance between easy and hard samples. 

The cross-entropy (CE) loss used to train the pitch prediction stage:
\begin{equation}
    \mathcal{L}_{\text{P}} = - \frac{1}{L} \sum_{i=1}^{L} \sum_{c=1}^{P} p_c^i \ln \left(  \frac{\exp (\hat{p}_c^i / T_2)}{ \sum_{k=1}^{C} \exp (\hat{p}_c^k / T_2)}  \right)
\end{equation}
where $T_2$ is the temperature hyperparameter.

\section{Details of Evaluation} \label{sec:ast-eval}

% \subsection{AST Evaluation} 

% Besides COn, COff, COnP, and COnPOff, we additionally apply average overlap ratio (AOR) and raw pitch accuracy (RPA) as objective metrics. 
% Given GT and corresponding predicted notes, their overlap ratio (OR) is defined as the ratio between the duration of the time segment in which the two notes overlap and the time segment spanned by the two notes combined. The AOR is given by the mean OR computed over all matching GT and predicted notes. 
% For RPA, we transform the GT and predicted note events into frame-level sequences and compute matching scores. Both metrics are implemented using the \texttt{mir\_eval} library \cite{raffel2014mir_eval}.

% For ROSVOT, we remove silence notes and designate the boundaries that enclose each note as onset and offset. This step is unnecessary for other baselines. The onset tolerance is set to 50 ms, and the offset tolerance is the larger value between 50 ms and 20\% of note duration. The pitch tolerance is set to 50 cents. All numbers demonstrated are multiplied by 100.

% \subsection{SVS Evaluation}

For each SVS experiment task, 20 samples are randomly selected from our test set for subjective evaluation. Professional listeners, totaling 20 individuals, are engaged to assess the performance. In MOS-Q evaluations, the focus is on overall synthesis quality, encompassing clarity and naturalness. For MOS-P, listeners are exposed to GT samples and instructed to concentrate on pitch reconstruction, disregarding audio quality. In both MOS-Q and MOS-P evaluations, participants rate various singing voice samples on a Likert scale from 1 to 5. It is crucial to highlight that all participants were remunerated for their time and effort, compensated at a rate of \$10 per hour, resulting in a total expenditure of approximately \$300 on participant compensation. Participants were duly informed that the data were for scientific research use. 

\section{Extensional Experiments} \label{sec:add}

\subsection{Additional AST Results}

The additional experimental results are listed in \autoref{tab:add1} and \autoref{tab:add2}, where the former is under a clean environment and the latter is noisy. 

\subsection{Additional Comparisons}

We perform additional out-of-domain (OOD) tests on ROSVOT and compare with \cite{gu2023deep}, which focuses more on multimodality, but still has noticeable audio-only AST capability. We directly run OOD tests on datasets MIR-ST500 \cite{wang2021preparation} and TONAS \cite{gomez2013towards}, where the former is used by \cite{gu2023deep} as an in-domain (ID) set and the latter also an OOD set. The ROSVOT model in this test is still trained with only M4Singer. The results are listed in \autoref{tab:comp} (the results of \cite{gu2023deep} are copied from their original paper).

\begin{table*}
\setlength\tabcolsep{10pt}
\centering
\setlength\tabcolsep{4pt}
\begin{tabular*}{0.67\hsize}{l|ccc|ccc}
\toprule
\multirow{2}{*}{ \bf Method }        & \multicolumn{3}{c|}{ \bf MIR-ST500 }    & \multicolumn{3}{c}{ \bf TONAS }  \\
                                     & COn   & COnP  & COnPOff                 & COn   & COnP  & COnPOff           \\
\midrule
\textit{\cite{gu2023deep}}           & 78.1  & 70.0  & 52.8                    & 64.4  & 36.9  & 24.1              \\
\textit{$\mathcal{M}$ (ours)}        & 72.1  & 65.9  & 47.4                    & 55.7  & 49.4  & 30.0           \\
\bottomrule
\end{tabular*}
\caption{\label{tab:comp}
Additional comparison with \cite{gu2023deep} on OOD tests.
}
\end{table*}

From the results, we can see that ROSVOT outperforms \cite{gu2023deep} in an OOD setting (i.e., on TONAS) in terms of COnPOff and COnP, but falls short in COn. We believe this is because \cite{gu2023deep} incorporated English data in training, which has a relatively closer pattern to flamenco singing than M4Singer. ROSVOT also underperforms facing MIR-ST500. There are three possible reasons:
\begin{itemize}[leftmargin=*]
    \setlength{\itemsep}{-4pt}
    \item Since each song in MIR-ST500 spans several minutes and needs to be source-separated first, it contains non-negligible un-voiced sections. We do not know whether \cite{gu2023deep} segments the voiced parts according to some rules. Hence, we transcribe the whole song and compute the scores at once, resulting in a smaller proportion of positive samples of on/offsets due to longer silences, hereby influencing the performance. Another noteworthy point is that ROSVOT is able to load a 5-minute song into one single 2080ti GPU and transcribe it at once, demonstrating considerable efficiency.
    \item The source-separation result influences the performance. The separator recognizes harmonies as vocals, thus resulting in polyphonic singing voices. We focus on solo vocal note transcription and harmonies are not considered noise in our settings. ROSVOT tends to transcribe vocal harmonies when the main vocal is silent, degrading the accuracy.
    \item MIR-ST500 is an ID test set to the model of \cite{gu2023deep}, so it should have an advantage.
\end{itemize}
In conclusion, we believe that ROSVOT demonstrates comparable or superior performance. Considering the total number of parameters, ROSVOT shows considerable efficiency and the capability to process long sequences. 

\subsection{Additional SVS Results}

In \autoref{tab:svs}, the quality of test samples could be very similar; without a specific evaluation objective, evaluators may struggle to determine differences in quality. Therefore, we conduct comparative evaluations and record CMOS scores of two typical data combinations compared to $100\%$ real $\mathcal{D}_1$. We believe CMOS scores, rating from -2 to 2, allow the disregard of unrelated factors. The results are listed in \autoref{tab:cmos}.

\begin{table*}
\setlength\tabcolsep{10pt}
\centering
\setlength{\belowcaptionskip}{-0.4cm}
\setlength\tabcolsep{5pt}
\begin{tabular*}{0.73\hsize}{cccc|cc|cc}
\toprule
\multirow{2}{*}{\bf Real}    & \multirow{2}{*}{\bf R. Size}  & \multirow{2}{*}{\bf Pseudo}       & \multirow{2}{*}{\bf P. Size}   & \multicolumn{2}{c|}{\bf CMOS-P}  & \multicolumn{2}{c}{\bf CMOS-Q}  \\
                             &                               &                                   &          & R & P & R & P     \\
\midrule
$\mathcal{D}_1\times10\%$  & 2.1    & $\mathcal{D}_1\times90\%$  & 18.8  & -0.26 & +0.08 & -0.34  & -0.05     \\
  -                        & 0.0    & $\mathcal{D}_1$            & 20.9  & -0.43 & +0.21 & -0.36  & +0.02     \\
\bottomrule
\end{tabular*}
\caption{\label{tab:cmos}
Results of comparative evaluation. Evaluators are requested to provide relative scores on each sample compared with reference samples which are generated using the SVS model trained with $100\%$ real $\mathcal{D}_1$.
}
\end{table*}

The results demonstrate a trend that, when the ratio of pseudo annotations in the training set increases, the quality of samples generated using real annotation inputs decreases, while that using pseudo annotations also increases. This may indicate that there are certain biases in the annotations within datasets, and the AST model could learn its own bias, which creates a domain gap of the annotation styles among different datasets. 

\section{Low-resource Scenarios} \label{sec:low}

\begin{figure}[htbp]
    \centering
    \includegraphics[width=0.45\textwidth]{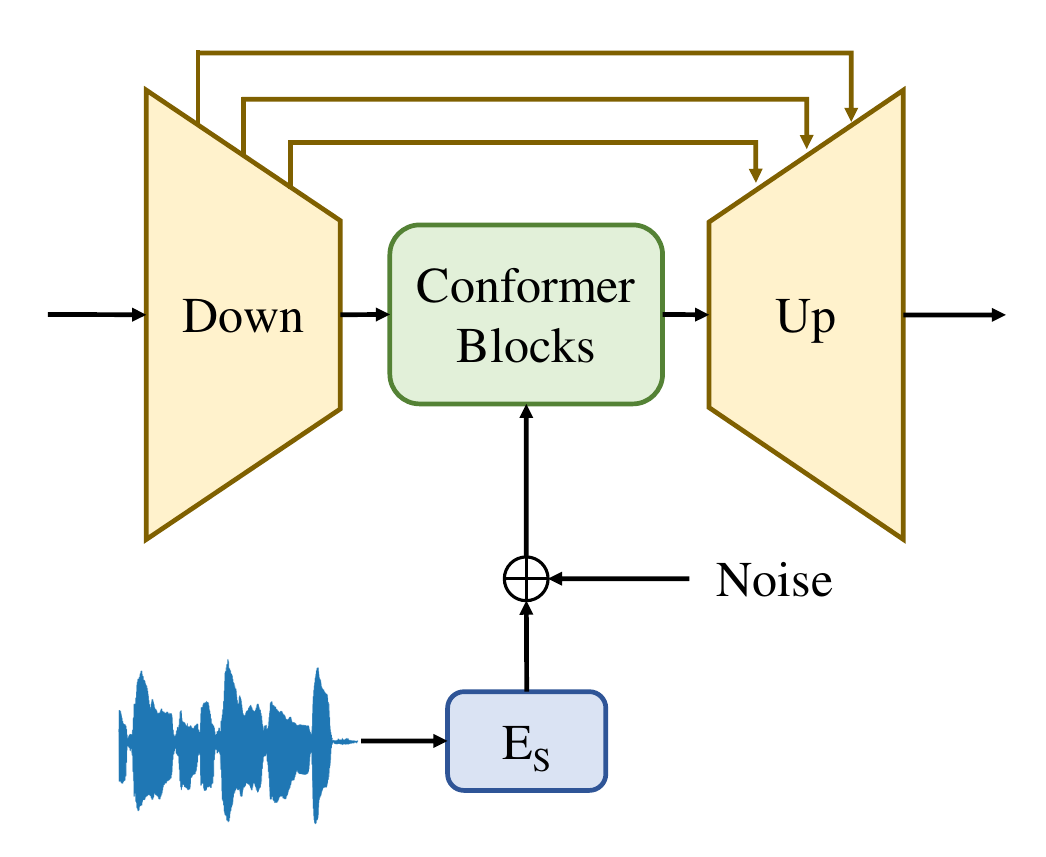}
    \setlength{\belowcaptionskip}{-0.5cm}
    \caption{
        Injection of self-supervised features.
    }
    \label{fig:ssl}
\end{figure}

\begin{table}[ht]
\setlength\tabcolsep{3pt}
\centering
\setlength{\belowcaptionskip}{-0.4cm}
\begin{tabular*}{\hsize}{cc|ccc}
% \begin{tabular*}{0.8\hsize}{@{}@{\extracolsep{\fill}}l|cc|c@{}}
\toprule
\bf Model                        & \bf Ratio  & \bf COn (F)     & \bf COff (F)      & \bf COnPOff (F)  \\
\midrule
$\mathcal{M}$                    & 100\%      &  93.7           & 94.3              & 77.1       \\
$\mathcal{M}$                    & 50\%       &  93.6           & 93.9              & 79.6       \\
$\mathcal{M}$                    & 10\%       &  93.0           & 92.6              & 71.6       \\
$\mathcal{M}$                    & 1\%        &  92.0           & 91.7              & 68.4       \\
\midrule[0.2pt]
\textit{$\mathcal{M}$(ssl)}      & 100\%      &  94.3           & 94.0              & 76.2       \\
\textit{$\mathcal{M}$(ssl)}      & 50\%       &  94.0           & 93.7              & 74.9       \\
\textit{$\mathcal{M}$(ssl)}      & 10\%       &  93.8           & 93.9              & 73.5       \\
\textit{$\mathcal{M}$(ssl)}      & 1\%        &  93.7           & 93.8              & 73.6       \\
\bottomrule
\end{tabular*}
\caption{\label{tab:low}
Results under low-resource scenarios. "Ratio" indicates the proportion of the training set that is utilized. 
}
\end{table}

Considering the scarcity of annotated singing voice datasets, we investigate the performance of the proposed method under low-resource scenarios. We use M4Singer as the training set and test the model on $\mathcal{D}_1$. Firstly, we gradually decrease the amount of training data to see the performance degradation. After that, we incorporate features extracted from a pre-trained self-supervised learning (SSL) framework to enhance the performance.

Specifically, we modify the model architecture by introducing a latent feature encoder $\text{E}_{\text{S}}$, transforming the additional SSL representations into 256-dimensional features, and performing a fusion by element-wise addition. This fusion can be illustrated as \autoref{fig:ssl}. $\text{E}_{\text{S}}$ comprises two convolution layers and a convolution block, where the former reduces the dimension of the input features to the model channel dimension. The output of $\text{E}_{\text{S}}$ is directly added to the output of the U-Net's encoder to perform the fusion. 

We choose XLSR-53 \cite{conneau2020unsupervised}, a wav2vec 2.0 \cite{baevski2020wav2vec} model pre-trained on 56k hours of speech in 53 languages, to be the SSL feature extractor. We believe that the knowledge of a pre-trained self-supervised model alleviates data scarcity. To simulate the low-resource environment, we actually can get access to singing voice corpora, only without annotations. Therefore, we use all the training data mentioned before to fine-tune the XLSR-53 model with a batch size of 1200k tokens for 20k steps. In this case, we incorporate self-supervised learning to cope with the low-resource problem. 

According to \citet{singla2022audio}, features from the second layer of a 12-layer wav2vec 2.0 model are the most related to audio features like pitch and unvoiced ratio, we extract features from the 4th layer of the 24-layer XLSR-53 to be the input feature, which has a dimension of 1024. Before feeding the features to the model, we add Gaussian noises with a standard deviation of 0.05 to perform the data augmentation. The SSL-augmented model is denoted as \textit{$\mathcal{M}$(ssl)}. 

The results are listed in \autoref{tab:low}. From the results, we can see that there is no significant improvement after involving SSL features, if enough training data is utilized. However, when decreasing the training data, the original model $\mathcal{M}$ exhibits a decline in performance, while \textit{$\mathcal{M}$(ssl)} experiences a comparatively smaller decrease.

\begin{table*}
\setlength\tabcolsep{10pt}
\centering
\setlength\tabcolsep{4pt}
\begin{tabular*}{0.98\hsize}{l|ccc|ccc|ccc|cccc}
\toprule
\multirow{2}{*}{ \bf Method }        & \multicolumn{3}{c|}{ \bf COn }    & \multicolumn{3}{c|}{ \bf COff }   & \multicolumn{3}{c|}{ \bf COnP }   & \multicolumn{4}{c}{ \bf COnPOff }   \\
                                     & P     & R     & F                 & P     & R     & F                 & P     & R     & F                 & P     & R     & F     & AOR           \\
\midrule
\textit{TONY}                        & 65.0  & 70.2  & 67.5              & 59.6  & 56.1  & 57.8              & 46.1  & 45.4  & 45.7              & 46.1  & 42.1  & 43.9  & 73.8         \\
\textit{VOCANO}                      & 73.7  & 78.1  & 75.8              & 69.1  & 73.5  & 71.2              & 56.3  & 59.9  & 58.0              & 48.4  & 52.1  & 50.2  & 81.4          \\
\textit{MusicYOLO}                   & 84.1  & 80.4  & 82.2              & 83.7  & 79.8  & 81.7              & 68.4  & 66.1  & 67.2              & 61.3  & 56.8  & 58.9  & 85.4         \\
\textit{\cite{yong2023phoneme}}      & 93.7  & 90.0  & 92.0              & 92.3  & 90.4  & 91.4              & 74.4  & 71.2  & 72.7              & 65.3  & 66.3  & 65.8  & 91.6         \\
\midrule[0.2pt]
\textit{$\mathcal{M}$ (conformer)}   & 90.2  & 93.0  & 93.7              & 92.3  & 96.0  & 94.1              & 78.2  & 80.0  & 79.1              & 75.3  & 77.4  & 76.3  & 97.0         \\
\textit{$\mathcal{M}$ (conv)}        & 92.2  & 91.0  & 91.6              & 92.8  & 92.3  & 92.6              & 73.1  & 72.1  & 72.6              & 71.6  & 70.2  & 70.9  & 96.8          \\
\textit{$\mathcal{M}$ (w/o noise)}   &\bf94.5& 93.2  & 93.8              &\bf94.7& 93.7  & 94.2              &\bf79.9& 79.3  & 79.6              & 76.0  & 76.8  & 76.4  &\bf97.1       \\
\textit{$\mathcal{M}$ (w/o wbd)}     & 92.1  & 94.5  & 93.3              & 91.6  & 94.8  & 93.2              & 79.2  & 80.9  & 80.1              & 75.9  & 78.4  & 77.1  & 96.5          \\
\midrule[0.2pt]
\textit{$\mathcal{M}$ (ours)}        & 92.4  &\bf96.2&\bf94.0            & 93.0  &\bf96.8&\bf94.5            & 79.5  &\bf81.2&\bf80.3            &\bf76.4&\bf78.9&\bf77.4& 97.0         \\
\bottomrule
\end{tabular*}
\caption{\label{tab:add1}
Additional results of AST systems. These results are from clean inputs.
}
\end{table*}

\begin{table*}
\setlength\tabcolsep{10pt}
\centering
\setlength\tabcolsep{4pt}
\begin{tabular*}{0.98\hsize}{l|ccc|ccc|ccc|cccc}
\toprule
\multirow{2}{*}{ \bf Method }        & \multicolumn{3}{c|}{ \bf COn }    & \multicolumn{3}{c|}{ \bf COff }   & \multicolumn{3}{c|}{ \bf COnP }   & \multicolumn{4}{c}{ \bf COnPOff }   \\
                                     & P     & R     & F                 & P     & R     & F                 & P     & R     & F                 & P     & R     & F     & AOR        \\
\midrule
\textit{TONY}                        & 48.6  & 49.8  & 49.2              & 45.9  & 48.2  & 47.0              & 32.3  & 36.4  & 34.2              & 25.8  & 31.7  & 28.4  & 46.6        \\
\textit{VOCANO}                      & 63.7  & 65.8  & 64.7              & 64.7  & 67.5  & 66.1              & 48.1  & 50.8  & 49.4              & 42.9  & 44.0  & 43.4  & 71.9        \\
\textit{MusicYOLO}                   & 81.3  & 78.2  & 79.7              & 79.6  & 73.7  & 76.5              & 62.1  & 59.2  & 60.6              & 53.9  & 49.9  & 51.5  & 78.6        \\
\textit{\cite{yong2023phoneme}}      & 89.7  & 87.2  & 88.5              & 90.1  & 89.3  & 89.7              & 69.2  & 66.7  & 67.9              & 63.4  & 60.8  & 62.1  & 86.4        \\
\midrule[0.2pt]
\textit{$\mathcal{M}$ (conformer)}   & 89.8  & 92.6  & 91.2              & 91.2  & 93.0  & 92.1              & 78.0  & 80.3  & 79.1              & 73.7  & 75.9  & 74.8  & 96.9       \\
\textit{$\mathcal{M}$ (conv)}        & 92.4  & 90.6  & 91.5              & 93.3  & 91.9  & 92.6              & 76.1  & 74.8  & 75.6              & 71.1  & 70.6  & 70.8  & 96.8        \\
\textit{$\mathcal{M}$ (w/o noise)}   & 91.9  & 89.9  & 90.9              & 92.3  & 90.7  & 91.5              & 78.8  & 77.4  & 78.1              & 70.2  & 70.0  & 70.1  & 95.2        \\
\textit{$\mathcal{M}$ (w/o wbd)}     & 92.4  & 94.6  & 93.5              & 91.6  & 94.2  & 92.9              & 83.2  & 83.8  & 83.5              & 76.3  &\bf77.7& 77.0  & 96.2        \\
\midrule[0.2pt]
\textit{$\mathcal{M}$ (ours)}        &\bf92.9&\bf94.7&\bf93.8            &\bf93.5&\bf95.3&\bf94.4            &\bf83.2&\bf84.4&\bf83.8            &\bf76.7& 77.3  &\bf77.0&\bf97.1      \\
\bottomrule
\end{tabular*}
\caption{\label{tab:add2}
Additional results of AST systems. These results are from noisy inputs.
}
\end{table*}

\end{document}